\documentclass[12pt]{article}
\pdfoutput=1
\usepackage[english]{babel}
\usepackage{amssymb,amsmath}
\usepackage{amsfonts}
\usepackage{latexsym}
\usepackage{verbatim}
\usepackage{amsthm}
\usepackage{amsbsy}
\usepackage{graphicx,color}
\usepackage{epsfig}
\usepackage{verbatim}
\usepackage{cite}
\usepackage[debug,pageanchor=false]{hyperref}
\definecolor{link}{rgb}{.8,.15,.1}
\hypersetup{colorlinks=true,linkcolor=link,citecolor=link,urlcolor=link,linktocpage}
\newcommand{\be}{\begin{equation}}
\newcommand{\ee}{\end{equation}}
\newcommand{\bi}{\begin{itemize}}
\newcommand{\ei}{\end{itemize}}
\newcommand{\bea}{\begin{eqnarray}}
\newcommand{\eea}{\end{eqnarray}}
\newcommand{\ba}{\begin{array}}
\newcommand{\ea}{\end{array}}

\def\ym2{{YM$_2$}}

\def\Tr{\mathrm{Tr}}
\def\rmd{{\rm d}}

\newcommand{\nn}{\nonumber}
\newlength{\sswidth}

\begin{document}
%%%%%%%%%%%% Title Page %%%%%%%%%%%%%%%%%%%%%%
\begin{titlepage}

\vskip 2cm
\begin{center}
%\def\thefootnote{\fnsymbol{footnote}}
%{\Large \textbf{Physical Matter, Topological Gravity}}
%\vskip0.45cm
%{\Large \textbf{and}}
%\vskip0.45cm
{\Large \textbf{New  Supersymmetric Localizations}} %backgrounds in two dimensions 
\vskip0.45cm
{\Large \textbf{from}}
\vskip0.45cm
{\Large \textbf{Topological Gravity}}
\end{center}
\vspace{1cm}

\vspace{1cm}

\begin{center}
  \textsc{~Jinbeom Bae$^{1,a}$,  ~Camillo Imbimbo$^{2,3,b}$, ~Soo-Jong Rey$^{1,4,c}$,~Dario Rosa$^{1,d}$}  
\end{center}

\vspace{0.5cm}

\begin{center}
\sl
 $^1$ School of Physics \& Astronomy and Center for Theoretical Physics\\
Seoul National University, Seoul 08826 \rm KOREA\\
\vspace{0.3 cm}
\sl
$^2\,$ Dipartimento di Fisica, Universit\`a di Genova,
Via Dodecaneso 33, 16146, Genoa, \rm ITALY \\
\sl $^3\,$ INFN, Sezione di Genova, Via Dodecaneso 33, 16146, Genoa, \rm ITALY\\
 \vspace{0.3 cm}
\sl
 $^4$ Fields, Gravity \& Strings, \ Center for Theoretical Physics of the Universe\\
Institute for Basic Sciences, Daejeon 34047 \rm KOREA
\vspace{0.3 cm}
\end{center}

\begin{center}
{\small $^a$kastalean4@gmail.com,~~$^b$camillo.imbimbo@ge.infn.it,  $^c$sjrey@snu.ac.kr,~~$^d$dario.rosa85@snu.ac.kr}
\end{center}

\vspace{1cm}

\centerline{\textsc{ Abstract}}
 \vspace{0.2cm}
 
{\small  
Supersymmetric field theories can be studied exactly on off-shell ``localizing'' supergravity backgrounds. 
We show that  these supergravity configurations can be identified  with BRST invariant configurations of {\sl background} topological gravity coupled to {\sl background} topological gauge multiplets.  We apply this topological point of view to two-dimensional ${\cal N}=(2,2)$ supersymmetric matter theories to obtain, in a simple and straightforward way, a complete classification of localizing supersymmetric backgrounds in two dimensions. We recover all known localizing backgrounds and (infinitely) many more that have not been explored so far.  The newly found localizing backgrounds are characterized by quantized fluxes for both graviphotons of the ${\cal N}=(2,2)$ supergravity multiplet. 
The BRST invariant topological backgrounds are parametrized by both Killing vectors and $\mathbb{S}^1$-equivariant cohomology of the two-dimensional spacetime. We completely reconstruct the supergravity backgrounds from the topological data: some of the supergravity fields are twisted versions of the topological backgrounds, but others are {\sl composite},  in that they are nonlinear functionals of topological fields.  Moreover,  we show that the supersymmetric $\Omega$-deformation is nothing
but the background value of the ghost-for-ghost of topological gravity, a result which  holds for higher dimensions too.}
\vspace{0.2cm}
\thispagestyle{empty}
%\end{center}

\vfill
\eject
\end{titlepage}
\hypersetup{pageanchor=true}
% to fix a bug with titlepage&hyperref: see comment after \usepackage[debug,pageanchor=false]{hyperref}
\baselineskip18pt
\tableofcontents
\vskip2cm
\rightline{\sl ``Fiyero....It's not lying. It's .... looking at things another way!''}
\rightline{Gregory Maguire}
\rightline{`Wicked: The Life and Times of the Wicked Witch of the West'} 
\newpage
\section{Introduction and Summary} 
\label{sec:intro}
The localization technique refers to an exact WKB method by virtue of which the semi-classical approximation becomes exact. 
It has been extensively studied for a broad class of quantum field theories that admit Lagrangian descriptions, in particular, supersymmetric or topological quantum field theories. 
For instance, the  topological quantum field theories (TQFTs) whose action is BRST-exact\footnote{These TQFTs are commonly known as ``of cohomological type''.} are semi-classically exact since their coupling constant is a gauge parameter which can be taken arbitrarily small. A traditional route for constructing TQFTs is by topologically twisting supersymmetric quantum field theories (SQFTs) by means of a conserved R-symmetry. Hence, the  localization technique has frequently been associated to SQFTs since the early days. 

Recently, starting from the work \cite{Pestun:2007rz}, localization technique has been revived for various SQFTs, without connecting it to TQFTs in any explicit manner. 
Rather, in this point of view,  localization is the consequence of the existence of  a global (nilpotent on physical states) supercharge
when the  SQFT is defined on specific external backgrounds.
These external backgrounds may be identified, as done first in \cite{Festuccia:2011ws}, 
with an off-shell configuration of a supergravity (SG) multiplet that the SQFT can couple to.
Global supercharges of the SQFT are in correspondence with 
generalized covariantly constant spinors that set the supersymmetry variations of fermionic fields of the SG 
multiplet to zero.

The generalized covariantly constant spinor must satisfy integrability conditions which put stringent constraints on the bosonic fields of the SG multiplet. 
These fields include the spacetime metric and also, in theories with extended supersymmetries, vector fields of gauged R-symmetries as well as off-shell auxiliary fields.   It turns out that, in general, the bosonic fields of the SG multiplet must be switched on for the SQFT to be put supersymmetrically on a  --- compact or noncompact --- curved manifold supporting generalized covariantly constant spinors. 
The background bosonic fields are external sources for associated conserved current operators of the SQFT, thus parametrize the space of deformed SQFT. Hereafter, we shall refer to this space of SQFTs as the {\it matter} SQFT. 

Generalized covariantly constant spinors  depend on the spacetime dimensions and also on the specific SG to which the matter SQFT is coupled. 
The localization technique to matter SQFT is applicable when the space of generalized covariantly 
constant spinors is non-empty. 
A complete classification of generalized covariantly constant spinors is a complicated problem. 
Although explicit solutions have been obtained case by case in various spacetime dimensions, 
there is no general strategy for constructing the covariantly constant spinors and for classifying 
the background spacetime metrics and gauge fields which support them.

In this paper, we put forward a new approach  
for finding localizing backgrounds for matter SQFTs. The strategy of our approach ---    which was already introduced by two of the authors of the
present paper  in the context of three-dimensional $\mathcal{N}=2$ supersymmetric gauge theories \cite{Imbimbo:2014pla} ---   is the following: one starts from a  SQFT and twists it to obtain a corresponding topological {\it matter}
theory which is then  coupled to  topological gravity (TG) {\it backgrounds}.  One then
seeks for BRST-invariant backgrounds: each BRST-invariant  background is associated with a topological matter theory.  The trivial background, of course, corresponds to the original topological matter theory. Non-trivial topological backgrounds define new topological matter theories whose deformations are
associated to the geometrical structures  parametrizing the BRST-invariant backgrounds. 

In the present paper, we apply this  topological approach to  two-dimensional matter SQFT and we elaborate on its general features.   We will see that  the equations for  covariantly constant spinors of  two-dimensional $\mathcal{N}=(2,2)$  supergravity are
recast in the topological framework  as  {\it cohomological} equations.  Specifically, we show  that the generalized covariantly constant spinors   are expressed in terms of $\mathbb{S}^1$-equivariant de Rham cohomology  of the underlying spacetime. The  cohomological formulation  incorporates
automatically the concept of {\it gauge equivalent} topological gravity backgrounds.  
We work out the explicit map between the BRST invariant topological backgrounds 
and the supersymmetry-preserving two-dimensional $\mathcal{N}=(2,2)$  supergravity backgrounds: this map 
allows therefore to  identify  supergravity backgrounds which are equivalent for localization purposes.  
To the best of our knowledge, the concept of {\sl equivalent} generalized covariantly constant spinors is new and appears to be one major benefit of our approach. We also explicitly construct all the inequivalent  localizing SG backgrounds. We recover the solutions found  in previous works \cite{Benini:2012ui} -\cite{Closset:2015rna}  and also many more new ones: in fact, infinitely many more. 

The equations for the BRST invariant topological backgrounds are the 
topological counterpart of the equations for the generalized covariantly constant spinors of SG. 
However, contrary to the na\" ive expectation,  the topological gravity system which is relevant in our framework is  {\it not} a  topological twist of the supergravity of the standard approach. 
In fact, the relation  that we uncover between the generalized covariantly constant spinors of the supergravity approach and the solutions of the cohomological equations of
the topological approach  is non-trivial.  
Most of topological bosonic background fields are not, in any sense, 
 bosonic fields of {\sl topologically twisted} SG.  Several of
the BRST invariant topological backgrounds are  bilinears 
of the  covariantly constant spinors of supergravity. For example, the ghost-for-ghost field\footnote{The ghost-for-ghosts of topological gauge  and gravity theories
are also referred to as  {\sl super-ghosts}.} of TG is identified with the spinorial bilinear which defines the Killing vector  of  the spacetime metric.
Conversely, the SG fields which correspond to a  BRST invariant topological background are, 
in general, non-linear functionals of the fields of TG. In this work, we explicitly construct
these functionals for the specific case of two-dimensional $\mathcal{N} = (2,2)$ SG.  

The construction we introduce, which is in essence based on general properties of Fierz identities, is in principle generalizable to higher dimensions and to higher extended supersymmetry.  However, the specific topological background systems to which one has to couple the topological matter  depend on the dimension and on the particular matter system one considers. For example, we found previously in \cite{Imbimbo:2014pla} that  the supersymmetric backgrounds of three-dimensional $\mathcal{N}=2$ supergravity were described by {\it pure} topological gravity. 
In the present paper we find instead that the topological description of localizations of two-dimensional $\mathcal{N}=(2,2)$ supergravity
requires, beyond topological gravity, also a topological background abelian multiplet. At the moment, we do not have yet an a priori way to identify the correct topological gravity backgrounds which describe localizations of a given matter supersymmetric system:  this remains an interesting open problem.  We should also add that in our approach we are restricted to the case with at least two supercharges,  which corresponds to backgrounds that satisfy a reality condition.   This case is the one for which the complex conjugate of the (generalized) covariantly constant spinor is also covariantly constant. This is sometimes referred to as the  {\sl real} case in the literature.  

The paper is organized as follows:

 In Section \ref{sec:2dYM}, we look for  a topological counterpart of  ${\cal N}=(2,2)$ supersymmetric gauge theory which  can be coupled to two-dimensional TG. To this end, we revisit the topological formulation of two-dimensional Yang-Mills (\ym2) theory  \cite{Witten:1992xu}~\footnote{For self-contained presentation, 
we recapitulate in Appendix \ref{App:AppendixA} connection between standard and topological \ym2 theories.}, which forms the vector multiplet part of the matter SQFTs\footnote{We do not describe in this paper the topological twist of supersymmetric chiral matter, since finding the correct coupling to topological gravity of the vector multiplet is sufficient for the goal of finding the localizing backgrounds.}.
We end up with a topological version of two-dimensional standard YM theory coupled to a  topological $U(1)$ field strength background, which can be alternatively thought of as a twisted version of two-dimensional $\mathcal{N}=(2,2)$ vector
multiplet. 

In  Section \ref{sec:superfield}, we find  the consistent coupling of this matter topological YM theory to background  TG. The resulting theory depends now on {\it two} topological backgrounds: the TG background and the topological $U(1)$ field strength background. We also identify the associated BRST transformations for both matter fields and backgrounds which close off-shell.

In Section \ref{sub:fixedpoints}, we classify the topological BRST invariant backgrounds.
As it is familiar from SG, the BRST invariant topological backgrounds are specified by the BRST transformations rules for the backgrounds only: they are independent of the specific matter TQFT which couples to them.  In the TG  approach  the
equations which specify the invariant backgrounds are obtained by setting to zero the BRST variations of the
two fermionic fields, i.e. the topological gravitino  $\psi^{\mu\nu}$  and gaugino $ \psi^{(1)} $.  
 The BRST variation of the topological
gravitino of TG  provides an equation for the metric and the bosonic ghost-for-ghost 
$\gamma^{\mu}$ of\ the TG multiplet
\bea
S\, \psi^{\mu\nu} = D^\mu \,\gamma^\nu + D^\nu \,\gamma^\mu =0. 
\label{equivariantKilling}
\eea
Simply put, these equations state that the ghost-for-ghost background is a Killing vector of the metric.
In two dimensions, non-trivial solutions of (\ref{equivariantKilling}) exist only if the euclidean spacetime  manifold  is either a  2-sphere $\mathbb{S}^2$ or a 2-torus $\mathbb{T}^2$, equipped with a metric possessing at least 
one isometry $V^\mu$.  Different topologies of the spacetime manifold  only
support the trivial solution $\gamma^\mu=0$ which corresponds to the Witten topological twist \cite{Witten:1993yc}.
The equation (\ref{equivariantKilling}) is universal, in the sense that holds for any topological gravity system,
in any dimension. We mentioned above that in two dimensions we must consider a topological gravity system which includes also a topological $U(1)$ multiplet background. Therefore, we obtain one more equation from the  BRST variation of the topological $U(1)$  background gaugino $ \psi^{(1)} $:
\bea
S\, \psi^{(1)} = \rmd \, \gamma^{(0)} - i_\gamma \ f^{(2)} =0, 
\label{equivariantS1onS2}
\eea
where $\gamma^{(0)}$ is the bosonic superghost of the $U(1)$ gauge multiplet background and $f^{(2)}$ is the $U(1)$ field strength. 
 
 Eq. (\ref{equivariantS1onS2}) is  the simplest and most extensively studied example of equivariant cohomology:
it states that the $U(1)$ topological backgrounds $( \gamma^{(0)}, f^{(2)})$ are  equivariant classes of the $\mathbb{S}^{1}$-equivariant cohomology
on  the $\mathbb{S}^2$ or $\mathbb{T}^2$ euclidean spacetime.  The interesting case is the one of the 2-sphere, $\mathbb{S}^2$. It is well-known that 
the $\mathbb{S}^1$-equivariant cohomology of the sphere is the polynomial ring generated by two classes $x$ and $y$ of ghost number 2,
subject to the hypersurface relation 
\bea
 x^2 -y^2=0 \ . 
 \label{localizationring}
\eea
 We describe these classes in detail in Section \ref{sub:fixedpoints}. They parametrize the moduli  space of inequivalent SG backgrounds that lead to supersymmetric localization. 

In Section \ref{sub:compSUGRA}, we explain  the map between  BRST invariant backgrounds and localizing backgrounds of $\mathcal{N} = (2,2)$ SG. 
This SG multiplet contains two graviphotons. 
The 2-form $f^{(2)}$ is identified with the field strength of one of them. 
The superghost fields --- both the vector $\gamma^\mu$ of TG and the scalar $\gamma^{(0)}$ of the topological $U(1)$ 
gauge multiplet ---  which solve (\ref{equivariantKilling}) and (\ref{equivariantS1onS2})  coincide with the independent bilinears of 
the covariantly constant spinors of SG. In  two dimensions, there is another scalar spinorial bilinear, $c_0$, which is determined by the 
independent bilinears by  means of a quadratic relation. This scalar bilinear turns out to be related, via an equation
which is identical in form to (\ref{equivariantS1onS2}), to the field strength $G^{(2)}$ of the second graviphoton: this second graviphoton of SG
is therefore a ``composite'' field in terms of the topological variables.  We will provide the explicit expression for the second
graviphoton field strength in  terms of the topological fields.  Finally, we will also write down  the field strength of the $U(1)_R$ --- 
the R-symmetry of the supersymmetric matter theory ---  in terms of the topological backgrounds. In this way, {\it all} the bosonic fields of the SG multiplet which support generalized covariantly constant spinors are reconstructed in terms of the topological backgrounds 
solving  (\ref{equivariantKilling}) and (\ref{equivariantS1onS2}).

In Section \ref{sec:classification}, we analyze in detail the case of the two-sphere $\mathbb{S}^2$\footnote{The torus is also described by our formulas, but since in this case the $\mathbb{S}^1$ acts without fixed points, the equivariant  cohomology does not give more information than  the standard one.}. We recover all the known localizing solutions
which have been described in the literature.
We also uncover an infinite number of new solutions.   The structure of the space of supersymmetric backgrounds is qualitatively different for vanishing and non-vanishing superghost background $\gamma^\mu(x)$.

When  $\gamma^\mu(x)=0$, our equations imply that the two graviphoton field strengths $f$ and $G$ are equal, $f=G$. We will see that this, in turn, forces the  $U(1)_R$ field strength  $\mathcal{F}_R$ to coincide with  half of the two-dimensional spacetime curvature $R$. These supersymmetric
backgrounds correspond therefore to the old $A$-twisted topological matter models introduced by
Witten long ago \cite{Witten:1993yc}. These backgrounds exists for any topology of the two-dimensional spacetime.

When  $\gamma^\mu(x)=\epsilon_\Omega\, V^\mu(x)\not= 0$, where $V^\mu(x)$ the isometry of the sphere and $\epsilon_\Omega$ is the degree-two generator of the ring of $\mathbb{S}^1$-equivariant cohomology,  the space of localizing backgrounds  acquires new branches. In Figure \ref{FluxesSphere} the supersymmetric solutions are labelled by the  quantized fluxes   $(n, m)$ of the two graviphotons field stregths $(f, G)$. The solutions with $f=G$ and thus with $\mathcal{F}_R=\pm \frac{1}{2}\, R$ have now necessarily
zero fluxes $n=m=0$. These solutions,  corresponding to the green dot of  Figure \ref{FluxesSphere},
are the $\Omega$-deformed sphere backgrounds of  \cite{Closset:2014pda} and \cite{Closset:2015rna}.

A non-vanishing superghost $\gamma^\mu(x)$ also allows  for solutions with $f\not = G$.   If $|n|=|m|$  the $U(1)_R$ flux is still $\pm 1$, but the spin connection cannot be identified with (twice) the $U(1)_R$ gauge field. 
These solutions, depicted in Figure \ref{FluxesSphere} with red and blue dots,  depend on a {\it continuous} parameter $A$  ---  the zero-mode of the gauge scalar superghost $\gamma^{(0)}$\footnote{In the models for which the matter vector multiplet includes
a quadratic twisted superpotential, this continuous parameter can be identified with its coupling constant.}. 

There is a second class of solutions with $|n|\not= |m|$,  for which the zero-mode of $\gamma^{(0)}$ is {\it discrete} since it is identified with (half of) the flux $m$  of the ``composite'' graviphoton. For these solutions $m$  takes integer values in the set $\{-n, -n+1,\ldots, n\}$.
These ``discrete'' solutions have $U(1)_R$ flux equal to zero.  They are the black dots of Figure \ref{FluxesSphere}. The solution with $n=-2$ and $m=0$ is the solution studied in  \cite{Benini:2012ui} and \cite{Doroud:2012xw}. 

 We emphasize that the deformation parameter $\epsilon_{\Omega}$ 
 is non-vanishing for all the solutions corresponding to the black, red and blu dots: in this sense
 we can say that these solutions all have non-trivial Omega-background  since,  in the topological gravity formulation,  {\it the natural definition of Omega-background is  the vector superghost $\gamma^\mu$ background.}
This definition includes the standard Omega-deformed $\mathbb{S}^2$ as a particular case (the green dot) when the graviphoton field strengths are equal to each other. But it is more general and it applies to any spacetime dimensions since the form of the gravitino BRST variation of TG (\ref{equivariantKilling}) is universal. For example, its relevance in three dimensions was discussed in \cite{Imbimbo:2014pla}. In four dimensions, Nekrasov's Omega-deformation of twisted $N=2$ super Yang Mills theory \cite{Nekrasov:2002qd} is also captured by the superghost of the corresponding topological gravity. 

One should keep in mind that, because of ghost-number conservation, a  non-trivial  dependence of the partition function on the Omega background comes about only if one
considers insertions of suitable operators carrying non-trivial ghost number. 
This poses an interesting and nontrivial lesson of our construction, applicable to any  dimensions: in topological models, it is natural ---  and necessary to describe the full set of localizable SQFTs ---  to switch on topological backgrounds with non-vanishing and {\it even} ghost number. 

In section \ref {sec:lorentzian}, we describe the action of the non-compact $SO(1,1)$ duality group of SG on
the  localizing backgrounds. This non-compact duality group acts on the central charges of the supersymmetry algebra and it is an automorphisms of the generalized covariantly constant spinor equations  \cite{Closset:2014pda}. In the topological framework,
the duality group is the group of linear automorphism of the ring relation (\ref{localizationring}) which characterizes the BRST invariant backgrounds. 
The duality group is {\it non-linearly} realized on the topological backgrounds but it acts linearly on the cohomology classes $x$ and $y$.  It is in general broken by a given localizing background; however,  {\it discrete} subsets of duality transformations map localizing
SG backgrounds to different ones. We describe explicitly  these discrete subsets for the various kind of localizing SG backgrounds in Section   \ref {sec:lorentzian}.  In section \ref {sec:conclusions}, we summarize our main results and discuss issues which may be worth of future investigations.

%%%%%%%%%%%%%%%%%%%%%%%%%%%%%%%%%%%%%%%%%%%%%%%%%%%%%%%%
\section{A Topological Formulation of $d=2$ Yang-Mills Theory} 
\label{sec:2dYM}

The bosonic sector of two-dimensional ${\cal N}=(2,2)$ supersymmetric gauge theory contains the \ym2 theory. 

In this section, we develop a topological formulation of  \ym2 theory, viewed as a deformation of topological \ym2 theory. 
We consider both theories defined on a smooth manifold $\Sigma$ equipped with a metric $g_{\mu\nu}$.  

The relation between standard \ym2 theory and topological \ym2 theory was investigated long ago in \cite{Witten:1992xu}.    
Witten's reformulation of \ym2 theory, although closely related to the topological theory, is not invariant under reparametrizations: it explicitly
depends on a two-dimensional metric $g_{\mu \nu}$.  Here, we revise Witten's formulation and obtain a matter TQFT which can be consistently coupled to TG. 

Let us  first review Witten's formulation of \ym2 theory, whose bosonic field content is identical to that of the topological counterpart:
it consists of the gauge connection one-form field  $A= A^a\, T^a $ 
and a scalar field $\phi=\phi^a\, T^a$, both transforming in 
the adjoint representation of the gauge group $G$. 
$T^a$, with $a=1,\ldots \mathrm{dim}\, G$, are generators of 
the Lie algebra associated to the gauge group $G$.
The theory's partition function is
\bea
Z[g, \epsilon] = \int [\rmd A \rmd \phi] \ e^{ - I_{\rm YM}(g, \epsilon)}. \nn
\eea
Here,  $I_{\rm YM}$ is the action functional
\bea
I_{\mathrm{YM}}(g, \epsilon) =  \int_{\Sigma} \Tr \,\phi \,F + \epsilon \int_{\Sigma} \rmd^2 x\,\sqrt{g}\,\frac{1}{2} \Tr\, \phi^2 \, 
\label{actionYM}
\eea
where $F$ is  the  field strength two-form
\bea
F = d\,A + A^2 \ . 
\nonumber
\eea
Note that the $\epsilon$-independent part of the action (\ref{actionYM}) coincides with the bosonic part of the action of the topological \ym2 theory. The partition function defines an effective action of the two-dimensional metric $g_{\mu \nu}$ and the deformation parameter $\epsilon$. 

The action (\ref{actionYM}) is invariant under the 
BRST gauge transformations $s$:
\bea
&& s_{\rm gauge}\, c=  - c^2 \ ,\nn\\
&& s_{\rm gauge}\, A = - D\, c \ ,\nn\\
&& s_{\rm gauge}\, \phi = -[c, \phi] \ .
\label{gaugeBRST}
\eea
The action $I_{\rm YM}$ is quadratic in the scalar field $\phi$ which can therefore be integrated out, 
yielding the physical Yang-Mills theory action:
\bea
I_{\mathrm{YM}} = -\frac{1}{\epsilon}\,  \int_{\Sigma} \rmd^2 x \sqrt{g} \,\frac{1}{2} \Tr \, F^2 \ .
\label{actionYMbis}
\eea
Recalling that $\epsilon$ was initially introduced as a parameter that 
deforms away from the topological gauge theory, we see that it is 
identifiable with the coupling constant of the physical gauge theory. 
Conversely, the classical limit of the physical gauge theory is the 
topological gauge theory. 

The theory (\ref{actionYM}) is neither invariant under diffeomorphisms nor under  conformal transformations of the two-dimensional spacetime $\Sigma(g)$. 
It is however invariant under  {\it area-preserving} diffeomorphisms, $w(\infty)$, which span an infinite-dimensional symmetry transformations. 
This huge {\sl global}  symmetry is the basis for anticipating exact solvability 
of the theory \cite{tHooft:1974hx}, \cite{Migdal:1976}.

To further study how the infinite-dimensional global symmetry constrains the theory, we promote it to a {\it local} symmetry by coupling the theory to suitable background fields. Accordingly, we replace the volume-form $ \sqrt{g} \ \rmd^2 x $  with a topological background given by a two-form field $f^{(2)}$. This changes the action (\ref{actionYM}) to
\bea
\widetilde{I} [f^{(2)}] = \int_{\Sigma} \Tr \,\phi \,F -\int_{\Sigma} f^{(2)} \frac{1}{2} \Tr\, \phi^2 \, \  .
\label{actionYMone}
\eea
This action does not have the same physical content as the original action.  By Hodge decomposition, the two-form field $f^{(2)}$ takes the form
\bea
f^{(2)} = \Omega^{(2)} + \rmd \, \Omega^{(1)} \ ,
\label{f2volumeform}
\eea
where $\Omega^{(1)}$ is a one-form and 
\bea
\Omega^{(2)} = \epsilon\,  \sqrt{g} \ \rmd^2 x \ ,
\label{gvolumeform}
\nn
\eea
is a  representative of $H^{2}(\Sigma)$.  

For  $\widetilde{I}[f^{(2)}]$ to be equivalent to  $I_{\mathrm{YM}}[g, \epsilon]$, one must remove the degrees of freedom associated
with $\Omega^{(1)}$. We can achieve this by letting the background $f^{(2)}$ transform under the BRST operator  as follows
\bea
s \, f^{(2)} = - \rmd \, \psi^{(1)} \ ,
\label{BRSTone}
\eea
where $\psi^{(1)}$ is a fermionic background of ghost number +1. However, 
the BRST transformation (\ref{BRSTone}) is degenerate:  we need therefore to further introduce a {\it ghost-for-ghost} background $\gamma^{(0)}$ of ghost number +2 
\bea
s \, \psi^{(1)} = - \rmd \, \gamma^{(0)} \ ,
\label{BRSTtwo}
\eea
with
\bea
s\,\gamma^{(0)}=0 \ .
\label{BRSTthree}
\eea
Since now the background field is not inert under $s$, the action (\ref{actionYMbis}) is no longer BRST invariant:
\bea
s \, \widetilde{I} =s\,\left(-\int_{\Sigma} f^{(2)}\,\frac{1}{2}\, \Tr\, \phi^2 \right)= \int_{\Sigma} \rmd\, \psi^{(1)}\,\frac{1}{2}\, \Tr\, \phi^2 =  -\int_{\Sigma} \psi^{(1)}\,\Tr\, \phi \, D\,\phi \ .
\nn
\eea
To restore the BRST invariance, one must modify the BRST transformation law for the connection one-form $A$ as
\bea
s \, A= -D\,c   +\psi^{(1)}\,\phi +\cdots \ .
\label{BRSTAdeformedone}
\eea
We see that the BRST variation of the first term in $\widetilde{I}$ cancels the BRST variation of the second term:
\bea
s \, \widetilde{I}= s\,   \int_{\Sigma} \Tr \,\phi \,F -\int_{\Sigma} \psi^{(1)}\,\Tr\, D\,\phi\, \phi=  - \int_{\Sigma} \Tr \,\phi\, D\,\bigl(\psi^{(1)}\,\phi)-\int_{\Sigma} \psi^{(1)}\,\Tr\, D\,\phi\, \phi=  0 \ .
\nonumber
\eea
The problem with the modified transformation (\ref{BRSTAdeformedone}) is that it is no longer nilpotent:
\bea
s^2\, A = -\rmd\,\gamma^{(0)}\,\phi+\cdots  \ .
\nn
\eea
To fix this, it is necessary to modify also the BRST transformation rule for the ghost field $c$
\bea
s\, c = - c^2+ \gamma^{(0)}\, \phi \ .
\nn
\eea
One finds that
\bea
s^2\,c =0 \ ,
\nn
\eea
Moreover, an extra term in $s^2\,A$ cancels the term proportional to $\rmd \,\gamma^{(0)}$:
\bea
s^2\, A = D\, \bigl(\gamma^{(0)}\,\phi\bigr) - \rmd \,\gamma^{(0)}\,\phi +\cdots = \gamma^{(0)}\, D\,\phi +\cdots \ .
\nn
\eea
Although this is still nonzero, the lack of nilpotency is now reduced to a term proportional to the equations of motion of $A$:
\bea
\frac{\delta \widetilde{I}}{\delta A} = D\,\phi \  = 0.
\nn
\eea
We conclude that, on-shell, the BRST transformation for $A$ is nilpotent:
\bea
s^2 \, A \simeq 0 \qquad \textrm{on-shell} \ .
\nn
\eea

There is a systematic method to extend the on-shell BRST invariance to off-shell \cite{Imbimbo:2014pla}. 
One starts by introducing a one-form field valued in adjoint representation of the gauge group $G$  
 \be
\widetilde{A}\equiv \tilde{A}^a_\mu\, T^a\, \rmd x^\mu \ ,
\nn
\ee
carrying ghost number -1. One also modifies the BRST transformation rule for $A$ by adding to it
a term depending on the newly introduced one-form field $\widetilde{A}$:
\bea
&& s \, A = - D\,c  + \psi^{(1)}\,\phi +\gamma^{(0)} \,\widetilde{A} \ . 
\label{BRSTA}
\eea
This modification makes the BRST operator $s$ nilpotent off-shell on all fields
\bea
s^2\, c= s^2\, A= s^2\,\phi= s^2\, \widetilde{A}=0 \qquad \textrm{off-shell} \ ,
\nn
\eea
assuming that  $\widetilde{A}$ transforms according to
\bea
&& s \, \widetilde{A}= - [c, \widetilde{A}] - D\, \phi \ .
\nn
\eea
The term proportional to $\gamma^{(0)}$ in (\ref{BRSTA}) spoils the BRST invariance of the action
\bea
s\, \widetilde{I} =  -\int_{\Sigma} \Tr \,\phi\, D\,\bigl(\gamma^{(0)}\,\widetilde{A})= \int_{\Sigma} \gamma^{(0)}\,\Tr \,D\,\phi\wedge \widetilde{A} \ .
\nn
\eea
One needs therefore to further modify the action by adding to it a term quadratic in $\widetilde{A}$,
\bea
I \equiv  \int_{\Sigma} \Tr \,\phi \,F - \int_{\Sigma} f^{(2)}\,\frac{1}{2}\, \Tr\, \phi^2 + \int_{\Sigma} \, \gamma^{(0)}\,\frac{1}{2}\, \Tr\, ( \widetilde{A}\wedge \widetilde{A}) \ , 
\label{Actionoffshell}
\eea
The final action $I$, which is still topological, is manifestly invariant
\bea
s\, I =0 \ ,
\nn
\eea
under BRST transformations acting on {\it both} the dynamical fields {\it and} the backgrounds
\bea
&& s\, c = - c^2+ \gamma^{(0)}\, \phi \ ,\nn\\
&& s\, A = - D\,c +\gamma^{(0)} \,\widetilde{A} + \psi^{(1)}\, \phi \ ,\nn\\
&& s\, \phi = -[c, \phi] \ ,\nn\\
&& s\, \widetilde{A}= - [c, \widetilde{A}] - D\, \phi \ ,\nn \\
&& s\, f^{(2)} = - \rmd \, \psi^{(1)} \ ,\nn\\
&& s\, \psi^{(1)} = - \rmd \, \gamma^{(0)} \ ,\nn\\
&& s\, \gamma^{(0)}=0 \ .
\label{BRSTfieldsbackgrounds}
\eea
Roughly speaking, we are introducing a set of spurion fields whose classical expectation values correspond to the backgrounds. 

To construct a theory invariant under {\it global}  topological supersymmetry,  we consider the backgrounds which are fixed points of the BRST transformations (\ref{BRSTfieldsbackgrounds}):
\bea\label{eq:constraintsnogravity}
\rmd\,\psi^{(1)}=0 \ \qquad \mbox{and} \qquad  \rmd\, \gamma^{(0)}=0 \ .
\nn
\eea
Given the $\mathbb{Z}_2$ grading structure of the ghost number, 
one can  choose the fixed point backgrounds to be purely {\it bosonic}: 
\bea
\psi^{(1)}=0\ \qquad \mbox{and} \qquad
\gamma^{(0)}= \textrm{constant} 
%\equiv \gamma^{(0)} 
 \ .
\label{fixedpoint}
\eea
In these backgrounds, the BRST transformations  act nontrivially  only on the dynamical fields
\bea
&& s\, c = - c^2+ \gamma^{(0)}\, \phi \ ,\nn\\
&& s\, A = - D\,c +\gamma^{(0)} \,\widetilde{A} \ ,\nn\\
&& s\, \phi = -[c, \phi] \ ,\nn\\
&& s\, \widetilde{A}= - [c, \widetilde{A}] - D\, \phi \ .
\label{BRSTfieldsrigid}
\eea
One can freely rescale the fields by the backgrounds to 
\bea
\widehat{\psi} \equiv \gamma^{(0)} \,\widetilde{A} \, \qquad  \mbox{and} \qquad
\widehat{\phi} \equiv  \gamma^{(0)}\, \phi \ , 
\nn
\eea
carrying ghost number 1 and 2, respectively. It is immediate to verify 
that the resulting theory is the quasi-topological Yang-Mills theory of  \cite{Witten:1992xu}. 

At this point, it would be helpful to recapitulate the strategy we have paved so far.  Our starting point is \ym2 theory, 
which has the area-preserving diffeomorphisms as an infinite-dimensional global symmetry. 
To implement this global symmetry systematically, we first replaced the two-dimensional volume form by a two-form spurion field $f^{(2)}$. 
The global symmetry of the original model is reflected by the fact that  the action containing the spurion fields depends only on the cohomology class of $f^{(2)}$.  
We showed that, as the background is promoted to a spurion field, this procedure entails both extending the BRST transformations 
to the spurion fields and accordingly deforming the BRST transformations of the gauge multiplet. This procedure amounts to promote the global symmetry to a local gauge symmetry. 
To ensure  off-shell BRST invariance  it was  necessary to introduce an anticommuting one-form field, $\widetilde{A}$, as a compensator. This field turned out to be proportional to the gaugino of topological \ym2.  We managed to obtain in this way a 
BRST invariant formulation of \ym2 theory coupled to topological spurion fields, viz. $(f^{(2)}, \psi^{(1)}, \gamma^{(0)})$.

A comment about the spurion topological multiplet $(f^{(2)}, \psi^{(1)}, \gamma^{(0)})$ is in order.  The role of $f^{(2)}$ was to replace the metric volume form $\sqrt{g}\, d^2x$ of the Witten formulation, as indicated in Eqs. (\ref{f2volumeform}) and (\ref{gvolumeform}). 
As such the topological spurion  multiplet is not necessarily to be identified with the field strength multiplet of a $U(1)$ topological connection.
However we will see that, in the correspondence that we will establish between topological backgrounds and localizing SG backgrounds,
$f^{(2)}$ will be identified with the field strength of one of the two SG  graviphotons. Therefore it is natural to require that
\bea
f^{(2)} = \rmd \, a^{(1)}
\eea
where $a^{(1)}$ is an abelian {\it connection} on $\Sigma$, which transforms under the BRST operator as follows:
\bea
s\, a^{(1)} = -\rmd \, \xi^{(0)}+ \psi^{(1)}\ . 
\eea
Here, $\xi^{(0)}$ is the ghost of the $U(1)$ gauge symmetry. 

The partition function
\bea
Z[(f^{(2)}, \psi^{(1)}, \gamma^{(0)})] \equiv  \int [\rmd A\, \rmd \phi\, \rmd \widetilde{A}] \  \exp \big(- I[ A, \phi,\widetilde{A}; f^{(2)}, \psi^{(1)}, \gamma^{(0)}] \big) \ ,
\nn
\eea
 which encodes the effective action of the spurion fields and hence the gauged global symmetry, satisfies the Ward identity
\bea
s \, Z[(f^{(2)}, \psi^{(1)}, \gamma^{(0)})]=0 \ . 
\eea
This identity expresses the fact that the partition function depends only on the cohomology class of $f^{(2)}$.  In the BRST formulation, this is the statement  that  \ym2 theory is invariant under the area-preserving diffeomorphisms, viz. the $w(\infty)$ algebra.

Theories invariant under the `rigid' topological supersymmetry are now obtained by restricting the spurion fields to the backgrounds which 
are {\it bosonic} fixed points of the deformed BRST operator, viz. $\gamma^{(0)}= $ constant and $\psi^{(1)}=0$, as explained in (\ref{fixedpoint}).  Hence,  there is a one-parameter 
family of theories, labelled by the BRST invariant constant background $\gamma^{(0)}$. Depending on the background value, the topological supersymmetry is realized differently. 

For non-degenerate backgrounds, $\gamma^{(0)}\not=0$, one recovers the topological \ym2 theory and also identifies the topological gaugino $\psi$, which remained mysterious in Witten's formulation \cite{Witten:1992xu}, with the ``composite'' spurion $\gamma^{(0)}\,\tilde{A}$. As the background is non-degenerate, the standard \ym2 theory has the topological supersymmetry as a manifest symmetry. 
For degenerate background, $\gamma^{(0)}=0$, one
recovers the original \ym2 theory (\ref{actionYM}). The topological supersymmetry collapses and the BRST symmetry reduces to the pure gauge one, (\ref{gaugeBRST}). Thus, when $\gamma^{(0)}=0$, the topological supersymmetry  can be thought of as a hidden symmetry of the standard Yang-Mills theory.

%%%%%%%%%%%%%%%%%%%%%%%%%%%%%%%%%%%%%%%%%%%%%%%%%%%%%%%%
\section{Coupling to $d=2$  Background Topological Gravity}
\label{sec:superfield}

We next couple the TQFT constructed in Section 2 to two-dimensional TG.  To this end, it  is useful to formulate the theory in terms of superfields (or polyforms).  

We introduce the dynamical superfields, both valued in the adjoint representation of the Lie algebra of the gauge group $G$:
\bea
&& \mathcal{A} \equiv c + A + \widetilde \phi \ , \nn \\
&& \Phi \equiv \phi + \widetilde{A} + \widetilde c \ , 
\nn
\eea
where $\widetilde \phi$ is a two-form of ghost number $-1$ and $\widetilde c$ is a two-form of ghost number $-2$. The 
 total fermion number is defined to be the sum of the form degree and the ghost number. So, $ \mathcal{A}$ carries fermion number $+1$, while $\Phi$ carries fermion number $0$. We also introduce the spurion superfield (whose expectation value yields the super-background)
 \bea
&& \mathbf{f}  \equiv \gamma^{(0)} + \psi^{(1)} + f^{(2)} \ ,
\nn
\eea
carrying total fermion number $+2$. One can show straightforwardly that the superfield relations
\bea
&& \delta_0\, \mathcal{A} + \mathcal{A}^2 = \mathbf{f} \, \Phi \ , \nn \\
&& \delta_0 \, \mathbf{f} = 0 \nn\\
&& \delta_0 \, \Phi+ [ \mathcal{A}, \Phi]=0,
\label{eq:vectorrigidpoly} 
\eea
where $\delta_0$ stands for the derivation
\bea
 \delta_0 \equiv s + \rmd \ ,
\nn
\eea
are equivalent to the BRST transformations (\ref{BRSTfieldsbackgrounds}) and also define the BRST transformations for  $\widetilde \phi $ and $ \widetilde c$:
\bea
&& s \, \widetilde \phi = - [c , \widetilde \phi] - F  + \gamma^{(0)} \tilde c + \psi^{(1)} \widetilde A + f^{(2)} \phi \ , \nn \\
&& s \, \widetilde c = - [c , \widetilde c] - [\widetilde \phi, \phi] - D \widetilde A \ . 
\nn
\eea
The BRST invariant action
\bea
{I} &=& \int_{\Sigma}\mathbf{f}\,\frac{1}{2}\, \mathrm{Tr}\,  \Phi^2 =\nonumber \\
&=& \int_{\Sigma}\, \left[ \,f^{(2)}\,\frac{1}{2}\, \mathrm{Tr}\,  \phi^2+ \psi^{(1)}\wedge \mathrm{Tr} \,\phi\, \widetilde{A}+   \gamma^{(0)}\, \mathrm{Tr}\,( \phi\, \tilde{c} + \frac{1}{2} \widetilde{A}\wedge \widetilde{A}) \right]
\label{eq:ActionSuperfields}
\eea
corresponds to the action (\ref{Actionoffshell}). To see this,  we solve for $\widetilde{\phi}$ and $\widetilde{c}$ by putting
\bea
\widetilde{\phi} =0 
\nn
\eea 
and 
\bea
s\, \widetilde{\phi} =- F+\gamma^{(0)} \widetilde c + \psi^{(1)} \tilde A + f^{(2)} \phi = 0 \ .
\nn
\eea
into the action (\ref{eq:ActionSuperfields}) and obtain the action (\ref{Actionoffshell}).

We are ultimately interested in putting the theory on curved spacetime and in a background with nontrivial gauge fields.  Therefore, we shall couple our topological formulation of \ym2 theory to two-dimensional TG. The field content of TG includes the metric $g_{\mu\nu}$,  the gravitino $\psi_{\mu\nu}$, the diffeomorphism ghost $\xi^\mu$, and  the ghost-for-ghost $\gamma^\mu$ needed for the nilpotency of the BRST charge. They carry  ghost numbers $0,1,1,2$, respectively. The BRST transformations  of these fields \cite{Baulieu:1988xs} are
\bea
\label{eq:topgravBRST}
	&& s \,g_{\mu\nu} = - \mathcal L_\xi g_{\mu\nu} + \psi_{\mu\nu} \ , \nn \\
&& s \,\xi^{\mu} = - \frac 12 \mathcal L_\xi \xi^\mu + \gamma^\mu \ , \nn \\
	&& s \, \psi_{\mu\nu} = - \mathcal L_\xi \psi_{\mu\nu} + \mathcal L_\gamma g_{\mu\nu} \ , \nn \\
&& s \, \gamma^\mu = - \mathcal L_\xi \gamma^\mu \ ,
\eea
where $\mathcal L_\xi$ is the Lie derivative associated with the vector field $\xi$. 

It is useful to introduce the operator $S$
\bea
 S \equiv s + \mathcal L_\xi \,
\nn
\eea
which  satisfies the relation
\bea
\label{eq:lgammasquare}
S^2 = \mathcal{L}_\gamma \
\eea
on all the fields except the vector field $\xi^\mu$. 
Finding a nilpotent BRST operator $s$ for the matter TQFT coupled to TG is equivalent to finding an operator $S$ that satisfies the relation (\ref{eq:lgammasquare}) on the matter sector.

The solution to this problem\cite{Imbimbo:2009dy} is obtained by replacing the coboundary operator $\delta_0$ with a new nilpotent operator $\delta$: 
\bea
\label{eq:coboundarydelta}
 \delta \equiv S + \rmd - i_\gamma \ = \delta_0 + \mathcal L_\xi - i_\gamma , \qquad \delta^2 =  0 \ 
\eea
in the transformations rules (\ref{eq:vectorrigidpoly}):
\bea
\label{eq:vectorcoupledpoly}
&& \delta \mathcal{A} + \mathcal{A}^2 =  \mathbf{f} \, \Phi \ , \nn \\
&& \delta\, \mathbf{f} = 0 \ .
\eea
The equations above describe the BRST transformation rules for topological \ym2 theory coupled to TG. In components, these transformations become
\bea
&&  S\, c = - c^2+ \gamma^{(0)}\, \phi  + i_\gamma A \ \nn\\
&& S\, A = - D\,c +\gamma^{(0)} \,\widetilde{A} + \psi^{(1)}\, \phi + i_\gamma \widetilde \phi \ \nn\\
&& S \, \widetilde \phi = - [c , \widetilde \phi] - F  + \gamma^{(0)} \tilde c + \psi^{(1)} \widetilde A + f^{(2)} \phi \  \nn \\
&&S\, \phi = -[c, \phi] + i_\gamma \widetilde A \ \nn\\
&& S\, \widetilde{A}= - [c, \widetilde{A}] - D\, \phi  + i_\gamma \widetilde c \ \nn \\
&& S \, \widetilde c = - [c , \widetilde c] - [\widetilde \phi, \phi] - D \widetilde A \nn \\
&& S\, f^{(2)} = -\rmd\, \psi^{(1)} \ \nn\\
&&S\, \psi^{(1)} = - \rmd\, \gamma^{(0)} + i_\gamma f^{(2)} \ \nn\\
&& S\, \gamma^{(0)}= i_\gamma \psi^{(1)} \ .
\label{BRSTfieldsbackgroundscoupled}
\eea

Most importantly, the action (\ref{eq:ActionSuperfields}) remains BRST invariant even when the spacetime manifold $\Sigma$ is curved. In general, we can add to the action terms of the form
\bea
I_n = \frac{a_n}{n}\,  \int_{\Sigma}\mathbf{f}\, \mathrm{Tr}\,  \Phi^n,  \qquad (n=2,\cdots) \ . 
\label{eq:ActionnSuperfields}
\eea
In particular, in case the gauge group $G$ contains $U(1)$ factors, we can add a topological counterpart of the Fayet-Iliopoulos term
\bea
I_1 = a_1\,  \int_{\Sigma}\mathbf{f}\, \mathrm{Tr}\,  \Phi=a_1 \int_{\Sigma}\bigl(f^{(2)}\, \mathrm{Tr}\,\phi +  \psi^{(1)}\wedge \mathrm{Tr}\,\tilde{A} +    \gamma^{(0)}\, \mathrm{Tr}\,\tilde{c}\bigr) \ , 
\label{eq:ActionFISuperfields}
\eea
which, after eliminating $\widetilde{\phi}$ and $\widetilde{c}$, becomes
\bea
I_1 \simeq a_1 \,  \int_{\Sigma}\, \mathrm{Tr}\, F  \ . 
\nn
\eea
For $n=2$, we regain the BRST invariant action (\ref{eq:ActionSuperfields}).

%%%%%%%%%%%%%%%%%%%%%%%%%%%%%%%%%%%%%%%%%%%%%%%%%%%%%%%%
\section{BRST invariant Topological Backgrounds}
\label{sub:fixedpoints}

Having coupled the matter TQFT to TG, we now look for the background configurations that are BRST invariant.  The  BRST invariance conditions for the fermionic fields of both TG and topological $U(1)$ multiplet read
\bea
S \,\psi_{\mu\nu} = 0  \qquad  \mbox{and} \qquad
S \,\psi^{(1)} =0.
\nn
\eea
They lead to the equations
\bea
&& \mathcal{L}_\gamma g^{\mu\nu}=D^\mu \,\gamma^\nu + D^\nu \,\gamma^\mu =0,\nn\\
&&\rmd \, \gamma^{(0)} = i_\gamma f^{(2)} \, , 
\label{fixedbackgr}
\eea
characterizing the backgrounds in correspondence of which the matter QFT acquires global topological supersymmetry. Our aim is to solve these equations and classify the solutions modulo BRST trivial ones.

The action depends on the topological backgrounds only through the BRST operator $S$. The BRST operator, in turn, depends on the ghost-for-ghost $\gamma^\mu$ of TG and on the $U(1)$ fields $\gamma^{(0)}$ and $f^{(2)}$ only. Therefore, when the equations (\ref{fixedbackgr}) are satisfied, the matter QFT is automatically independent of any variation of the metric that preserve $\gamma^\mu$, as well as of any topological
variation of the $U(1)$ fields that preserve the class of $f^{(2)}$. 

The first equation in (\ref{fixedbackgr}) asserts that the ghost-for-ghost $\gamma^\mu$ has to be a Killing vector of the two-dimensional metric $g_{\mu\nu}$. This equation takes the same form in any spacetime dimensions, but the moduli space of solutions differs. In the context of three-dimensional supersymmetric gauge theories, its moduli space was discussed in \cite{Imbimbo:2014pla}. In two dimensions, it is well-known that there are Killing vectors on the sphere $\mathbb{S}^2$ and on the torus $\mathbb{T}^2$, but not on higher-genus Riemann surfaces. Given the Killing vector $\gamma^\mu$, we conclude that the matter QFT is independent of any $\gamma$-invariant  deformations of the metric.  Generically, these metrics have only a $U(1)$ isometry.  This is the case for example for the squashed two-sphere $\mathbb{S}^2_q$ studied in \cite{Gomis:2012wy} and \cite{Closset:2014pda}.

Given the Killing vector $\gamma^\mu$, we now turn to study the equations for the $U(1)$ gauge field background. 
The consistency of (\ref{fixedbackgr}) requires 
\bea
&&\rmd \, i_\gamma f^{(2)}=0= \mathcal{L}_\gamma\, f^{(2)}\nn\\
&& i_\gamma\, \rmd\gamma^{(0)} =0 =\mathcal{L}_\gamma \gamma^{(0)} \ . 
\nn
\eea
This means that all the backgrounds $g_{\mu\nu}$,  $f^{(2)}$ and $\gamma^{(0)}$ must be $\mathcal{L}_\gamma$-invariant. Trivial solutions are of the form
\bea
\gamma^{(0)} = i_\gamma (\theta^{(1)}) \qquad \mbox{and} \qquad f^{(2)}= \rmd \theta^{(1)} \ , 
\nn
\eea
where $\theta^{(1)}$ is a globally defined $\mathcal{L}_\gamma$-invariant 1-form:
\bea
\mathcal{L}_\gamma\, \theta^{(1)}=0 \ . 
\nn
\eea
The second equation in  (\ref{fixedbackgr}) 
\bea
\rmd \, \gamma^{(0)} - i_\gamma f^{(2)} = 0
\label{S1equivariant}
\eea
has the form of the defining equation of the equivariant closed form of degree-two of the $\mathbb{S}^1$-equivariant cohomology on a two-surface:
\bea
(\rmd - \epsilon_{\Omega}\,i_V)\, (f^{(2)} + \epsilon_{\Omega}\,f^{(0)})=0 \ , 
\label{CartanS1equivariant}
\eea
provided we make the identifications 
\bea
\gamma^{(0)} = \epsilon_{\Omega} \, f^{(0)}\qquad 
\mbox{and} \qquad \gamma^\mu = \epsilon_{\Omega} \, V^\mu
\nn
\eea
where $V$ is the Killing vector associated with the $\mathbb{S}^1$-equivariant action and $\epsilon_{\Omega}$ is the degree-two generator of the 
ring of the $\mathbb{S}^1$-equivariant cohomology.  The  $f^{(2)} + \epsilon_{\Omega}\,f^{(0)}$, which appears in (\ref{CartanS1equivariant}), is  the equivariantly closed extension of the ordinary differential form $f^{(2)}$ and $\rmd- \epsilon_{\Omega}\,i_V$ is the Cartan differential. 

 For $\Sigma = \mathbb{S}^2$,  it is well-known that there are two linearly independent equivariant classes $x$ and $y$ of degree-two~\footnote{We present an elementary proof of this assertion in the Appendix \ref{app:sphereS1S2}.}. The first class is the ring variable itself:
\bea
x =\epsilon_{\Omega}.
\nn
\eea
The second class is
\bea
y = \widetilde{f}^{(2)} + \epsilon_{\Omega}\, \tilde{f}^{(0)} \ , 
\nn
\eea
where
\bea
&& \widetilde{f}^{(2)}= \sqrt{g}\,\frac{1}{2} \epsilon_{\mu\nu} \rmd x^\mu\, \rmd x^\nu \nn\\
&&  D^2\, \widetilde{f}^{(0)}=\sqrt{g} \,\epsilon_{\mu\nu}\, D^\mu \, V^\nu \ .  
\nn
\eea
Here, $\widetilde{f}^{(0)}$ is solved only up to an additive constant: given a choice of this constant, a shift to another value induces the change 
 \bea
 y \quad \to \quad y + c\,x \ . 
\nn
 \eea
We choose the normalization of  the variable $y$ such that
\bea
\int_{\Sigma} y=\int_{\Sigma}  \widetilde{f}^{(2)} =-2 \qquad \mbox{and}  \qquad y(N) = \epsilon_{\Omega}\,\widetilde{f}^{(0)}(N)=  \epsilon_{\Omega} \, 
\label{equivnorm1}
\eea
where $N$ is one of the fixed points of the vector field $V$. If $S$ is the other fixed point of $V$, we can choose
\bea
-2= \int_{\Sigma}  \widetilde{f}^{(2)} = \widetilde{f}^{(0)}(S)- \tilde{f}^{(0)}(N)\qquad \mbox{and} \qquad  \widetilde{f}^{(0)}(S)= - \widetilde{f}^{(0)}(N)=-1 \ . 
\label{equivnorm2}
\eea
The localizing SG background found in \cite{Benini:2012ui} corresponds to an equivariant class of the form $ a\, y + b \, x$ with $a\not=0$,
whereas the background  identified in \cite{Closset:2015rna} corresponds to a class with $a=0$. 

The square of $y$ is an equivariantly closed class of degree-four:
\bea
y^2 = x\, ( 2\,\widetilde{f}^{(2)} \,   \widetilde{f}^{(0)}+ \epsilon_{\Omega}\, (\tilde{f}^{(0)})^2) \ . 
\nn
\eea
Hence, we have
\bea
D \, ( 2\,\widetilde{f}^{(2)} \,   \widetilde{f}^{(0)}+ \epsilon_{\Omega}\, (\widetilde{f}^{(0)})^2)=0 \ , 
\nn
\eea
from which we derive the cohomological equation
\bea
 2\,\widetilde{f}^{(2)} \,   \widetilde{f}^{(0)}+ \epsilon_{\Omega}\, (\widetilde{f}^{(0)})^2\quad \sim \quad  \alpha\, x+ \beta\, y \ . 
\nn
\eea
Here, $\alpha$ and $\beta$ are determined by
\bea
&& \int_\Sigma y^2 =  \epsilon_{\Omega}\,\int_\Sigma  2\,\widetilde{f}^{(2)} \,   \widetilde{f}^{(0)}=  \beta\, \int_{\Sigma} x\, y= \beta\,\epsilon_{\Omega}\nn\\
&& \nn \\
&& y^2(N) = \epsilon_{\Omega}^2 \, (\widetilde{f}^{(0)}(N))^2 =\alpha\, x(N)^2 + \beta\, x(N)\, y(N) = (\alpha + \beta)\,\epsilon_{\Omega}^2 \ . 
\nn
\eea
This yields
\bea
\beta = \int_\Sigma  2\,\widetilde{f}^{(2)} \,   \widetilde{f}^{(0)}\qquad \mbox{and} \qquad \alpha = -\beta + 1 \ . 
\nn
\eea
With the normalizations (\ref{equivnorm1}) and (\ref{equivnorm2}), we have
\bea
\beta=0 \ .
\nn
\eea
We thus obtain the cohomological relation
\bea
y^2 \sim x^2 \ ,
 \label{cohomologyrelation}
\eea
which tells us that the $\mathbb{S}^1$-equivariant cohomology at any degree is the polynomial ring generated by $x$ and $y$ modulo the relation (\ref{cohomologyrelation}).

Throughout the above analysis, we were taking both the TG  backgrounds and the $U(1)$ field strength $f^{(2)}$ background to be {\it real-valued}. This implies that, for $\Sigma$ a compact surface, the flux of  $f^{(2)}$ must be quantized. Hence, on the two-sphere, the relevant cohomology is the integer valued $\mathbb{S}^1$-equivariant cohomology. 
In Section \ref{subsec:sphere}, we will discuss the impact of this quantization condition on the topological moduli space. 

For $\Sigma = \mathbb{T}^2$, the $\mathbb{S}^1$-action is free. So, the equivariant cohomology is the same as the standard cohomology of the quotient $\mathbb{T}^2/\mathbb{S}^1\simeq \mathbb{S}^1$~\footnote{See appendix \ref{subsec:torus} for an explicit verification of this well-known general statement regarding $\mathbb{T}^2$.}. As such, there is just one parameter for the inequivalent BRST invariant topological backgrounds.

%%%%%%%%%%%%%%%%%%%%%%%%%%%%%%%%%%%%%%%%%%%%%%%%%%%%%%%%
\section{Relation to Supergravity Backgrounds}
\label{sub:compSUGRA}

Given the classification of the topological backgrounds just discussed, our next goal is to establish a map between the  topological BRST invariant backgrounds and the supersymmetric backgrounds of two-dimensional $\mathcal{N} = (2,2)$ SG. 
We will show that the equations determining the BRST invariant topological backgrounds which we derived in the previous Section are equivalent to the equations for the generalized covariantly constant spinors of $\mathcal{N} = (2,2)$ SG in two dimensions.
It will be clear from our discussion that the method we will explain is very general, and it can be applied to other dimensions or to higher supersymmetry contents. We expect that the topological system which describe localizing backgrounds of  SG with higher supersymmetry and/or in higher dimensions  will include more topological gauge multiplets beyond the single
abelian one which we considered in this paper. 

The localizing backgrounds of SG are determined by the generalized Killing spinor equation, 
which is obtained by requiring the vanishing of the supersymmetry variations of the gravitino \cite{Closset:2014pda}:
\bea
(D_\mu - i \mathcal{A}_\mu) \,\zeta = - \frac{1}{2} \,H \,\Gamma_\mu \zeta + \frac{i}{2}\, G \,\Gamma_\mu \Gamma_3 \zeta \ . 
\label{eq:killingspinor}
\eea
Here, the covariant derivative $D_\mu$ includes the spin connection associated with the frame rotation on the tangent space $T\Sigma$, 
the vector field  $\mathcal{A}_\mu$ is the $U(1)_R$ gauge field  minimally coupled to the R-symmetry current, and the scalar fields $H$ and $G$ are the Hodge duals of the two graviphoton field strengths. The cases with at least two supercharges of opposite R-charges are the ones discussed in the previous works since they lead to amenable computations. They correspond to backgrounds in (\ref{eq:killingspinor}) which satisfy the following reality conditions
\bea
\mathcal{A}_\mu^* = \mathcal{A}_\mu \ ,  \qquad H^* = -H \ ,  \qquad G^* =  G \ . 
\label{eq:realbkg}
\eea
For these backgrounds, the conjugate of  (\ref{eq:killingspinor}) reads
\bea
(D_\mu + i \mathcal{A}_\mu)\, \zeta^\dagger = + \frac{1}{2}\, H \, \zeta^{\dagger} \Gamma_\mu - \frac{i}{2}\, G \,  \zeta^\dagger \Gamma_3 \Gamma_\mu \ . 
\label{eq:killingspinorconjugate}
\eea

%%%%%%%%%%%%%%%%%%%%%%%%%%%%%%%%%%%%%%%%%%%%%%%%%%%%%%
\subsection{Graviphoton Backgrounds}

The map between the TG backgrounds and the SG backgrounds is obtained by considering the decomposition of the bi-spinor in two dimensions:
\bea
\zeta_a(x)\, \zeta_b^\dagger(x) = c_0(x) \, \frac 1 2 \mathbb{I}_{ab} + c_\mu(x) \frac 1 2 \Gamma^\mu_{ab} + \widetilde{c}_0(x)\,\frac 1 2  \Gamma^3_{ab} \ , 
\nn
\eea
where
\bea
c_0(x) = \zeta^\dagger(x)\, \zeta(x)\ , \qquad c_\mu(x) = \zeta^\dagger(x)\Gamma_\mu \zeta(x)\ , \qquad \widetilde{c}_0(x)= \zeta^\dagger(x)\,\Gamma^3\, \zeta(x)
. 
\nn
\eea
The Fierz identities in two dimensions lead to the relation  
\bea
&& %||c_m||^2(x) := c_m (x) \, g^{mn} (x) \, c_n (x) =
c^\mu c_\mu = 
%\zeta^\dagger_a\, \Gamma^\mu_{ab}\, \zeta_b\,   \zeta^\dagger_c\, (\Gamma_\mu)_{cd}\, \zeta_d\nn\\
%&&\qquad = \frac 1 4 \Gamma^\mu_{ab}\,( c_0(x) \,  \mathbb{I}_{bc} + c_\nu(x)\, \Gamma^\nu_{bc} + \tilde{c}_0(x)\, \Gamma^3_{bc})\times\nn\\
%&&\qquad \times(\Gamma_\mu)_{cd}\,(c_0(x) \, \mathbb{I}_{da} + c_\rho(x) \, \Gamma^\rho_{da} + \tilde{c}_0(x)\,  \Gamma^3_{da})=\nn\\
%&&\qquad = \frac 1 4 \mathrm{Tr} \bigl[ (\Gamma^\mu\,c_0(x)  + c_\nu(x)\, \Gamma^\mu\,\Gamma^\nu + \tilde{c}_0(x)\, \Gamma^\mu\,\Gamma^3)\times\nn\\
%&&\qquad \times(\Gamma_\mu\, c_0(x)  + c_\rho(x) \, \Gamma_\mu \,\Gamma^\rho+ \tilde{c}_0(x)\,  \Gamma_\mu\Gamma^3)\bigr]=\nn\\
 c_0^2(x) - \widetilde{c}_0^2(x),
\label{fierz2d}
\eea 
where  we raised the indices of $c_\mu$ using the background metric $g^{\mu\nu}$.
The equations (\ref{eq:killingspinor}) and (\ref{eq:killingspinorconjugate}) for the spinors $\zeta$ and $\zeta^\dagger$ imply the following equations for the bilinears $c_0, c_\mu, \widetilde{c}_0$
\bea
\label{eq:bilinears}
&&D_\mu \, c_\nu +D_\nu \, c_\mu =0\nn\\
&&D_\mu \, c_\nu =  \sqrt{g}\,\epsilon_{\mu\nu} \, (G\, c_0 +i\, H\, \widetilde{c}_0)\nn\\
&&D_\mu \, \widetilde{c}_0=   -i\, H\,\sqrt{g}\,\epsilon_{\mu\nu}\,c^\nu \nn\\
&& D_\mu \, c_0=  G\,\sqrt{g}\,\epsilon_{\mu\nu} \,c^\nu \ , 
\eea
where we used the relation
\bea
\zeta^\dagger\, \Gamma_\mu \,\Gamma^3\, \zeta= -i\,\sqrt{g}\,\epsilon_{\mu\nu}\, \zeta^\dagger\,\Gamma^\nu \,\zeta=-i\,\sqrt{g}\,\epsilon_{\mu\nu}\,c^\nu \ . 
\nn
\eea

The SG variables $H$, $c^\mu$, and $\widetilde{c}_0$ have to be identified with the topological background fields according to the following map:
\bea
f \equiv \ast f^{(2)}= - i\, H,  \qquad \gamma^{(0)} = \widetilde{c}_0, \qquad \gamma^\mu = c^\mu \ . 
\label{dictionaryone}
\eea
Therefore, in correspondence to  a solution of the topological equations
\bea
D^\mu\, \gamma^\nu + D^\nu \, \gamma^\mu =0 \qquad \rmd \,\gamma^{(0)} = i_\gamma (f^{(2)}) \ , 
\label{topgravitybkgs}
\eea
we can construct a  solution of the equations (\ref{eq:bilinears}),  which is defined by  
\bea
&& c_0 = \sqrt{\gamma^2 + (\gamma^{(0)})^2}\nn\\
&& G=\frac {1}{c_0} \left[ { \frac{1}{2}\,\sqrt{g}\,\epsilon_{\mu\nu}\, D^\mu\,\gamma^\nu + f\, \gamma^{(0)}} \right].
%{ \sqrt{\gamma^2 +(\gamma^{(0)})^2}}
\label{dictionarytwo}
\eea
together with Eqs. (\ref{dictionaryone}).

As explained in the previous Section, solutions of the topological equations (\ref{topgravitybkgs}) that are related by the transformations
\bea
f^{(2)} \to f^{(2)}+ \rmd\, \omega^{(1)} \ ,  \qquad \gamma^{(0)} \to \gamma^{(0)} + i_\gamma(\omega^{(1)}) \qquad \mbox{where} \qquad \mathcal{L}_\gamma \,\omega^{(1)}=0
\label{topbkgrequivalence}
\eea
with  globally defined $\omega^{(1)}$,  are {\it gauge equivalent}. The flux of $f^{(2)}$ is, by definition, invariant under the gauge transformations (\ref{topbkgrequivalence}). Let us see if the same is true for the flux  of the SG background $G$. Under the
gauge transformations (\ref{topbkgrequivalence}),  the associated composite two-form field
\bea
G^{(2)} =  G \,\sqrt{g} \, \rmd^2 x =  \frac{1}{c_0} \left[ \frac{1}{2}\, \rmd\,k + f^{(2)}\, \gamma^{(0)} \right] \ , 
%{\sqrt{\gamma^2 +(\gamma^{(0)})^2}}
\nn
\eea
varies by  
\bea
\delta G^{(2)} &=& \frac{1}{c_0} \, \Big[ \rmd \,\omega^{(1)}\, \gamma^{(0)} + f^{(2)}\, i_\gamma(\omega^{(1)}) \Big]- \frac{G^{(2)}}{c_0}\, \frac{\gamma^{(0)}\, i_\gamma(\omega^{(1)})}{c_0} \nn \\
&=&
%\nn\\
% &&\qquad = \frac{1}{c_0} \,d\,(\omega^{(1)}\, \gamma^{(0)})- \frac{G^{(2)}}{c_0}\, \frac{\gamma^{(0)}\, i_\gamma(\omega^{(1)})}{c_0}=\nn\\
% &&\qquad = d\,( \frac{\omega^{(1)}\, \gamma^{(0)}}{c_0})+ \frac{d\, c_0}{c_0^2}\, \omega^{(1)}\, \gamma^{(0)}- \frac{G^{(2)}}{c_0^2}\, \gamma^{(0)}\, i_\gamma(\omega^{(1)})=\nn\\
%&&\qquad=
 \rmd \,( \frac{\omega^{(1)}\, \gamma^{(0)}}{c_0})\equiv \rmd\, \widetilde{\omega}^{(1)} \ , 
\nn
\eea
where
\bea
\widetilde{\omega}^{(1)} \equiv \frac{\gamma^{(0)}}{\sqrt{\gamma^2 +(\gamma^{(0)})^2}}\, \omega^{(1)}\qquad \mbox{and} \qquad \mathcal{L}_\gamma \,\widetilde{\omega}^{(1)} =0 \ . 
\nn
\eea
Moreover,
\bea
\delta\,  c_0 = i_\gamma(\widetilde{\omega}^{(1)}) \ . 
\nn
\eea
Hence we conclude that, under the gauge transformations (\ref{topbkgrequivalence}), the ``composite'' fields $G^{(2)}$ and $c_0$ transform in the same way as the topological fields $f^{(2)}$ and $\gamma^{(0)}$:
\bea
 G^{(2)} \to G^{(2)}+ \rmd \, \widetilde{\omega}^{(1)} \ ,  \qquad c_0 \to c_0+ i_\gamma(\tilde{\omega}^{(1)})  \qquad \mbox{where} \qquad \mathcal{L}_\gamma \,\widetilde{\omega}^{(1)}=0 \ . 
\label{dualtopbkgrequivalence}
\eea
In particular,
\bea
\int_{\Sigma} G^{(2)} =\int_{\Sigma}  \frac{ \frac{1}{2}\,\rmd\,k + f^{(2)}\, \gamma^{(0)}}{\sqrt{\gamma^2 +(\gamma^{(0)})^2}}\qquad \mbox{where} \qquad k \equiv g_{\mu\nu}\, \gamma^\nu\, \rmd x^\mu
\nn
\eea
is invariant under the BRST transformations, provided that the backgrounds satisfy the BRST invariance equations (\ref{topgravitybkgs}).

%%%%%%%%%%%%%%%%%%%%%%%%%%%%%%%%%%%%%%%%%%%%%%%%%%%%%
\subsection{$U(1)_R$ Field Strength Background}

We have seen how the topological backgrounds specify the spinorial bilinears and thus the backgrounds $H$ and $G$.
Below, we show how the $U(1)_R$ field strength is also obtained from the same topological backgrounds.

From the equation for the generalized covariantly constant spinors: 
\bea
(D_\mu - i\, \mathcal{A}_\mu) \, \zeta= \frac{i}{2}\, (-f\, \Gamma_\mu + G\, \Gamma_\mu\, \Gamma_3)\,\zeta
\ , 
\label{covariantlyconstantsugra}
\eea
we obtain
\bea
(D_\nu - i\, \mathcal{A}_\nu)\,(D_\mu - i\,\mathcal{A}_\mu) \, \zeta &=& \frac{i}{2}\, ( -(D_\nu\,f)\, \Gamma_\mu +(D_\nu\, G)\, \Gamma_\mu \, \Gamma_3)\,\zeta + \nn\\
&+& \frac{i}{2}\, (-f\, \Gamma_\mu  +G\, \Gamma_\mu\, \Gamma_3)\, \frac{i}{2}\, (-f\, \Gamma_\nu + G\, \Gamma_\nu\, \Gamma_3)\,\zeta \nn\\
&=& \frac{i}{2}\, \Bigl[ -(D_\nu\,f)\, \Gamma_\mu +(D_\nu \, G)\, \Gamma_\mu \, \Gamma_3 +  \frac{i}{2}\, (f^2   - G^2)\, \Gamma_\mu \,\Gamma_\nu)\Bigr]\,\zeta \ . 
\nn
\eea
Antisymmetrizing with respect to $\mu, \nu$ yields
 %\footnote{$[\Gamma_\mu, \Gamma_\nu]= 2\,i\,\sqrt{g}\,\epsilon_{\mu\nu}\, \Gamma_3$}
\bea
&&\!\!\!\!\!\!\!\!\!\!\left( - \frac{i}{2}\, \sqrt{g}\,R\,\epsilon_{\mu\nu}\, \Gamma_3+ i \,\mathcal{F}_{\mu\nu} \right) \, \zeta=  \frac{i}{2}\, \Bigl[(D_{[\mu}\,f)\, \Gamma_{\nu]} - (D_{[\mu}\, G)\, \Gamma_{\nu]}\, \Gamma_3 - \sqrt{g}\,\epsilon_{\mu\nu}\,(f^2   - G^2)\, \Gamma_3\Bigr]\,\zeta
\ ,  \nonumber
\eea
where
\bea
&& \mathcal{F}_{\mu\nu}= \partial_\mu\, \mathcal{A}_\nu - \partial_\nu \, \mathcal{A}_\mu \equiv\mathcal{F}_R\, \sqrt{g}\,\epsilon_{\mu\nu}\nn\\
&& [D_\mu, D_\nu] \,\zeta = \sqrt{g}\, \epsilon_{\mu\nu}\, R\, \frac{i}{2} \Gamma_3\, \zeta \ . 
\nn
\eea
Hence, we arrive at the equation
%\footnote{$-i\,\Gamma^\nu\, \Gamma_3 = \frac{\epsilon^{\mu\nu}}{\sqrt{g}}\, \Gamma_\mu$}
%\bea
%&&( -R\, \Gamma_3+ 2\,\mathcal{F}_R )\, \zeta=  \Bigl[ - \frac{\epsilon^{\mu\nu}}{\sqrt{g}}\, (D_{\nu}\,f)\, \Gamma_{\mu} + \frac{\epsilon^{\mu\nu}}{\sqrt{g}}\, (D_{\nu}\, G)\, \Gamma_{\mu}\, \Gamma_3 - (f^2   - G^2)\, \Gamma_3\Bigr]\,\zeta
% =\nn\\
%&&\qquad = \Bigl[ i\,\Gamma^\mu\, \Gamma_3 (D_{\mu}\,f) -i\,\Gamma^\mu\, (D_{\mu}\, G) - (f^2   - G^2)\, \Gamma_3\Bigr]\,\zeta 
%\eea
%That is
\bea
&&\bigl[ 2\,\mathcal{F}_R +(f^2- G^2-R)\, \Gamma_3+ (\sqrt{g}\,\epsilon_{\mu\nu}\,D^{\nu}f +i\,D_{\mu}\, G)\,\Gamma^\mu \, \bigr]\,\zeta =0 \ .  
\label{ARbkgr}
\eea
Nontrivial solutions of this equations exist whenever
%For supersymmetry-preserving backgrounds we have to impose that the null space is nontrivial. This requirement leads to the condition:
\bea
\det \bigl[ 2\,\mathcal{F}_R +(f^2- G^2-R)\, \Gamma_3+ (\sqrt{g}\,\epsilon_{\mu\nu}\,D^{\nu}f +i\,D_{\mu}\, G)\,\Gamma^\mu\, \bigr]=0 \ , 
\label{integrabilityU1R}
\eea
that is, 
\bea
2\,\mathcal{F}_R = \pm \sqrt{(f^2- G^2-R)^2 +D_\mu\,f D^\mu f - D^\mu G\,D_\mu \, G + 2\, i\, \sqrt{g}\,\epsilon_{\mu\nu}\, D^\mu\, G\, D^\nu \,f} \, . 
\label{integrabilityU1Rbis}
\eea
For generic $f$ and $G$, (\ref{integrabilityU1Rbis}) would  require that $\mathcal{F}_R$ be {\it complex-valued}. 
However, since the fluxes are annihilated by the Lie derivative along  $\gamma$:
\bea
\mathcal{L}_\gamma\, f= \mathcal{L}_\gamma\, G=0 \, 
\nn
\eea
it follows that\footnote{This can be proven as follows.  $\gamma^\mu\, D_\mu\, f=0$ implies 
\bea
\gamma^2 \,D_\mu \, f = (\gamma^2\, D_\mu - \gamma_\mu \,\gamma^\rho\, D_\rho)\, f= \sqrt{g}\, \epsilon_{\mu\nu}\,\gamma^\nu \, (\epsilon^{\rho\sigma}\,\gamma_\sigma\, D_\rho\,f).
\nn
\eea
So, $D_\mu f= \sqrt{g}\, \epsilon_{\mu\nu}\,\gamma^\nu \, A(f)$ where $A(f)\equiv  \frac{\epsilon^{\rho\sigma}\,\gamma_\sigma \, D_\rho\,f}{\gamma^2}$. The same holds for $G$. It then follows immediately that $\sqrt{g}\,\epsilon_{\mu\nu}\, D^\mu \, G\, D^\nu \,f=0$.  }
\bea
\sqrt{g}\,\epsilon_{\mu\nu}\, D^\mu\, G\, D^\nu\,f=0
\nn
\eea
and thus the square of the field strength, $\mathcal{F}_R^2$,  is {\it real-valued}:
\bea
\mathcal{F}_R = \pm \frac{1}{2}\,\sqrt{(f^2- G^2-R)^2 +D_\mu \,f D^\mu f - D^\mu G\,D_\mu \, G} \ . 
\label{FRfinal}
\eea
As a matter of facts, not only the square but also $\mathcal{F}_R$  itself is {\it real}. 
%We will check this explicitly in the next Section.  
This can be understood as follows. We first rewrite the integrability condition (\ref{integrabilityU1R}) in a different form. From (\ref{ARbkgr}), we have
\bea
 && 2\,\mathcal{F}_R\, c_0 +(f^2- G^2-R)\, \gamma^{(0)}+ (\sqrt{g}\,\epsilon_{\mu\nu}\,D^{\nu}f +i\,D_{\mu}\, G)\,\gamma^\mu=0\nn\\
 &&  2\,\mathcal{F}_R\, \gamma^{(0)} +(f^2- G^2-R)\, c_0+ (\sqrt{g}\,\epsilon_{\mu\nu}\,D^{\nu}f +i\,D_{\mu}\, G)\,i\,\frac{\epsilon^{\mu\rho}}{\sqrt{g}}\,\gamma_\rho=0
\ . 
\nn
 \eea
Since $f$ and $G$ are $\mathcal{L}_\gamma$ invariant, the imaginary terms in the equations above drop out:
 \bea
 && 2\,\mathcal{F}_R\, c_0 +(f^2- G^2-R)\, \gamma^{(0)}- D^{\mu}f \,\sqrt{g}\,\epsilon_{\mu\nu}\,\gamma^\nu=0\nn\\
 &&  2\,\mathcal{F}_R\, \gamma^{(0)} +(f^2- G^2-R)\, c_0- D^{\mu}\, G\,\sqrt{g}\,\epsilon_{\mu\nu}\,\gamma^\nu =0 \, . 
\label{integrabilitybkgrs}
 \eea
Combining the two equations, we obtain a manifestly real-valued expression of the $U(1)_R$ field strength
\bea
\mathcal{F}_R &=&  ( D^{\mu}f \, c_0- \gamma^{(0)}\,D^{\mu}\, G)\,\frac{\sqrt{g}\,\epsilon_{\mu\nu}\,\gamma^\nu}{2\,\gamma^2} \nn\\
&=&  \frac{\epsilon^{\mu\nu}}{\sqrt{g}}\, D_\mu \,\Bigl[\frac{(f\, c_0 - G\, \gamma^{(0)})\,\gamma_\nu}{2\, \gamma^2}\Bigr] \ . 
\nn
\eea
The flux of $ \mathcal{F}_R$  does not necessarily vanish since  the vector field 
\bea
\mathcal{A}^\mu =\,\frac{1}{2}\, (f\, c_0 - G\, \gamma^{(0)})\,\frac{\gamma^\mu}{\gamma^2}
\nn
\eea
becomes singular at the zeros of the vector field $\gamma^\mu$.  If we perform the transformation (\ref{topbkgrequivalence}) on $f$ and $\gamma^{(0)}$, the
field strength  $ \mathcal{F}_R$ changes by a globally defined total derivative:
\bea
\mathcal{A}^\mu \to  \mathcal{A}^\mu + \bigl(\frac{\epsilon^{\rho\sigma}\,\partial_\rho\,\omega_\sigma}{2\, c_0\sqrt{g}}-\frac{\gamma^\rho\, \omega_\rho\, G}{2\,c_0^2}\bigr)\gamma^\mu \, , 
\nn
\eea
which implies that the flux of $\mathcal{F}_R$ is invariant under topological transformations.

From (\ref{integrabilitybkgrs}), we can also express the scalar spinorial bilinears in terms of the SG backgrounds:
\bea
\gamma^{(0)} &=& \frac{\sqrt{g}\,\epsilon_{\mu\nu}\,\gamma^\mu}{(D f)^2 - (D G)^2 } \Bigl[  (D^\nu f)(f^2- G^2-R)-(D^{\nu} G)\sqrt{(f^2- G^2-R)^2 +(D f)^2 - (D G)^2 } \Bigr]\nn\\
c_0 &=&  \frac{\sqrt{g}\,\epsilon_{\mu\nu}\,\gamma^\mu}{(Df)^2 - (D G)^2}\Bigl[ (D^\nu G) (f^2- G^2-R)- (D^{\nu} f) \sqrt{(f^2- G^2-R)^2 + (Df)^2 - (DG)^2}\Bigr] . 
\nn
\eea
\bigskip

We note that the field strength background (\ref{FRfinal}) encompasses all known backgrounds discussed in \cite{Closset:2014pda} as special cases. When (\ref{FRfinal}) is satisfied, the matrix in (\ref{ARbkgr}) is generically of rank-one. In this case,  the system has only {\it two} global supercharges.  The system has {\it four} global supercharges when the matrix has rank-zero, that is when the $U(1)_R$ field strength vanishes,
\bea
f^2- G^2-R = D_\mu \,f = D_\mu \, G=0 \qquad \mbox{and hence} \qquad \mathcal{F}_R=  0\ , 
\label{rank0}
\eea
which agrees with the results of \cite{Closset:2014pda}. 

Let us also observe that Eq. (\ref{FRfinal}) implies that whenever 
\bea
f = G
\label{Amodelcondition}
\eea
one has
\bea
\mathcal{F}_R = \pm  \frac{1}{2}\, R
\eea
We might take this as the definition of the $A$-model, whose twisting was indeed originally characterized by identifying the spin-connection with twice the $U(1)_R$ gauge field. 

From Eq. (\ref{dictionarytwo}) we see that $\gamma^\mu=0$ automatically implies the A-model condition
(\ref{Amodelcondition}): the corresponding backgrounds ---  i.e.  $\gamma^\mu=0$, $f=G$ and $\mathcal{F}_R = \pm  \frac{1}{2}\, R$ --- identify the old A-model introduced by Witten in  \cite{Witten:1993yc}. When instead
the A-model condition (\ref{Amodelcondition}) is satisfied by  $\gamma^\mu\not =0$ one obtains the so-called \cite{Closset:2014pda}  $\Omega$-deformed A-model on the sphere. We will verify this in detail in subsection \ref{subsec:omegadeformed}.

%%%%%%%%%%%%%%%%%%%%%%%%%%%%%%%%%%%%%%%%%%%%%%%%%%%%%%%%%%
\section{Classification of Supergravity Backgrounds}
\label{sec:classification}
Our considerations in earlier sections apply to  any two-dimensional spacetime $\Sigma$ equipped with a metric  which has an isometry.  In this Section we shall focus separately on $\Sigma = \mathbb{S}^2$ and $\mathbb{T}^2$. While non-compact $\Sigma = \mathbb{H}^2$ 
%and its compact quotient 
is an equally interesting case, due to new features, we shall relegate its study to a separate work.  As we explained in the previous sections, 
there is no loss of generality in taking the metrics on  $ \mathbb{S}^2$ and $\mathbb{T}^2$ be maximally  symmetric.

%%%%%%%%%%%%%%%%%%%%%%%%%%%%%%%%%%%%%%%%%%%%%%%%%%%%%%%%%
\subsection{All Supersymmetric Localizing  Backgrounds on $\mathbb{S}^2$}
\label{subsec:sphere}

Consider the round two-sphere $\mathbb{S}^2$ with coordinates 
\bea
\rmd s^2 = \rmd \theta^2 + \sin^2\theta\, \rmd \phi^2. 
\label{rounds2}
\eea
We take $\gamma^\mu$ to be proportional to one of the three Killing vector fields,
\bea
\gamma = \epsilon_{\Omega}\, \partial_\phi \equiv \epsilon_{\Omega}\, V \ . 
\eea 
Up to topological equivalences, we know that the general solution of (\ref{topgravitybkgs}) is given by
\bea
\gamma^{(0)} = \epsilon_{\Omega}\,\bigl(A- \frac{n}{2}\, \cos\theta\bigr) \qquad \mbox{and} \qquad f =\frac{n}{2}, 
\label{newbkgr}
\eea
where $A$ is a constant and $n \in \mathbb{Z}$ labels the first Chern class of the topological connection
\bea
\int_{\Sigma} \sqrt{g}\, f =\frac{\gamma^{(0)}(\pi) - \gamma^{(0)}(0)}{\epsilon_{\Omega}}= n \ . 
\eea
The modulus $A$ for a generic solution which is topologically  equivalent to (\ref{newbkgr}) can be expressed as
\bea
A = \frac{ \gamma^{(0)}(\pi) + \gamma^{(0)}(0)}{2\,\epsilon_{\Omega}} = \frac{n}{2} + \frac{\gamma^{(0)}(0)}{\epsilon_{\Omega}}  \ . 
\eea
This expression is topologically invariant since the Killing vector $\gamma^\mu$ vanishes at the poles of $\mathbb{S}^2$.

For the graviphoton background $G$, we obtain
\bea
G_{n,A} = \frac{ \frac{n}{2}\, A + (1- \frac{n^2}{4})\, \cos\theta}{\sqrt{\sin^2\theta +(A-\frac{n}{2}\,\cos\theta)^2}} \ , 
\eea
whose flux takes the value
\bea
m\equiv \int_{\Sigma} \sqrt{g}\, G_{n,A} = \frac{c_0(\pi) - c_0(0)}{\epsilon_{\Omega}} = \left\{ \begin{matrix} + n & \quad \mathrm{for} & \quad A  \ge + \frac{|n|}{2}\\
2\,A\,\text{sign}(n) &\quad \mathrm{for} & \quad |A| < \frac{|n|}{2}\\ 
- n & \quad \mathrm{for} & \quad A \le -\frac{|n|}{2}\end{matrix} \right.  \ . 
\eea
Therefore, by requiring the quantization of this flux, we see that, when $ |A| \ge\frac{|n|}{2}$, $A$ is a {\it continuous} moduli parameter of this family of solutions. On the other hand, when $|A|< \frac{|n|}{2}$, the quantization of the flux for $G$ imposes that $A$ be a {\it discrete} parameter, taking the $2\,n-1$ values
\bea
A = A(m) = \frac{m}{2}  \qquad \mathrm{for} \qquad m = -(n-1),\cdots, 0 , \cdots,  n-1
\eea
and
\bea
G_{n,m} = \frac{ \frac{n}{2}\,  \frac{m}{2} + (1- \frac{n^2}{4})\, \cos\theta}{\sqrt{\sin^2\theta +( \frac{m}{2}-\frac{n}{2}\,\cos\theta)^2}} \ . 
\eea

\begin{figure}
\vskip-1cm
 \centering
    \includegraphics[width=0.7\textwidth]{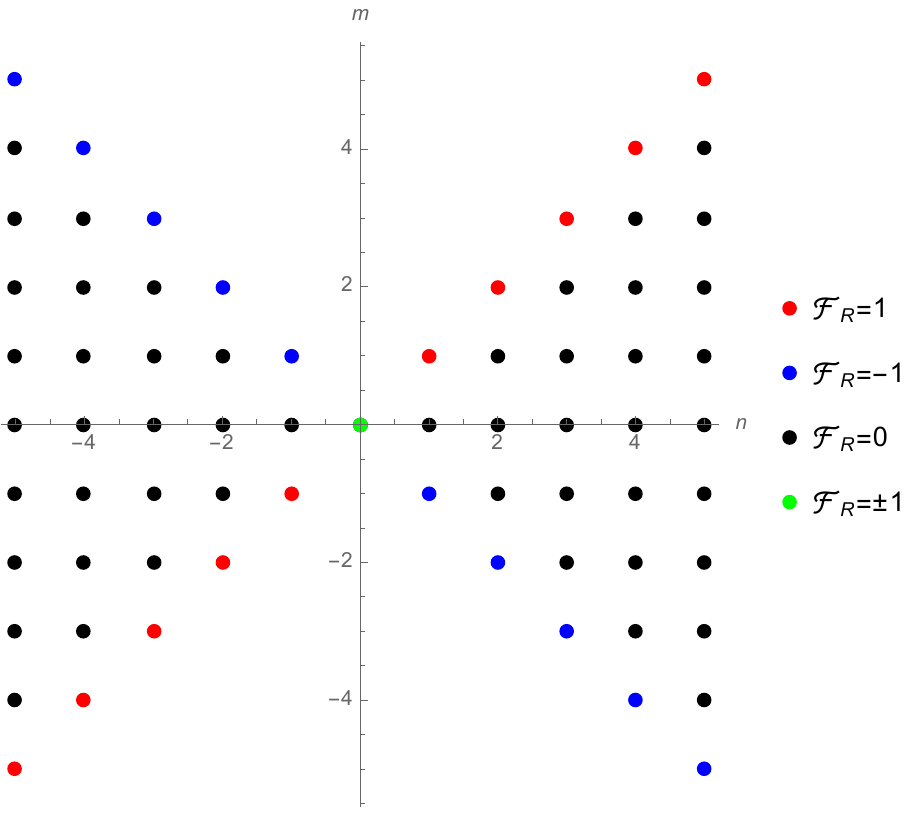}
   \caption{\sl Fluxes of localizing backgrounds on the 2-sphere: $n=\int_{\Sigma} \sqrt{g}\, f$, $m=\int_{\Sigma} \sqrt{g}\, G$.}
    \label{FluxesSphere}
\vskip0cm
\end{figure}

The $U(1)_R$ field strengths corresponding to the topological backgrounds (\ref{newbkgr}) are
\bea
\mathcal{F}^{\,n,A}_R = \pm \frac{1}{2}\,\sqrt{(n^2/4-1-G_{n,A}^2)^2  - (D G_{n,A})^2} \ . 
\eea
The flux of $U(1)_R$ gauge field is then given by 
\bea
\int_{\Sigma} \sqrt{g} \mathcal{F}^{\,n,A}_R\, &=& \int_0^\pi \rmd\theta \,\frac{\rmd}{\rmd \theta}\,\left[ (f\, c_0 - G\, \gamma^{(0)}) \frac{\gamma_{\phi}}{2\,\gamma^2} \right] \nn\\
&=& \frac{1}{2\,\epsilon_{\Omega}}\,\Bigl[(f(\pi)\, c_0(\pi) - G(\pi)\, \gamma^{(0)}(\pi))-(f(0)\, c_0(0) - G(0)\, \gamma^{(0)}(0))\Bigr]\nn\\
% &&\qquad  =\frac{1}{2}\,\Bigl[\frac{n}{2}\, |A + \frac{n}{2}| - \frac{\frac{n}{2}\, A + \frac{n^2}{4}-1}{|A+ \frac{n}{2}|}\, (A+\frac{n}{2})-(\frac{n}{2}\,  |A-\frac{n}{2}| - \frac{\frac{n}{2}\, A - \frac{n^2}{4}+1}{|A- \frac{n}{2}|}\, (A-\frac{n}{2}))\Bigr]=\nn\\
%&&\qquad =\frac{1}{2}\,\Bigl[ \frac{n}{2}\, ( |A + \frac{n}{2}| -|A-\frac{n}{2}|)+ \nn\\
%&&\qquad +\text{sign}(A -\frac{n}{2})\, (\frac{n}{2}\, A - \frac{n^2}{4}+1)-\text{sign}(A +\frac{n}{2})\, (\frac{n}{2}\, A + \frac{n^2}{4}-1)\Bigr]=\nn\\
&=& \left\{ \begin{matrix} +1 & \qquad \mathrm{for} & \qquad \ A \ge +\frac{|n|}{2}\\
\ 0 &\qquad \mathrm{for} & \qquad |A| < \frac{|n|}{2}\\ 
- 1 & \qquad \mathrm{for} & \qquad \ A \le -\frac{|n|}{2}
\end{matrix} \right. \ .  
\eea
In Figure \ref{FluxesSphere},  continuous and discrete solutions are represented on the $(m,n)$ plane, where  $m$ and $n$ are the $G$ and  $f$
fluxes. 
%we classified the supersymmetry-preserving backgrounds into continuous and discrete series. Here, the two-dimensional lattice points $(m,n)$ labels the integer-valued ($G$, $f$) fluxes.  
The solutions with continuous $A$ are pictured by red (blue) dots  on the lines $n=m$ ($n=-m$) and  their $U(1)_R$ flux is $+1$ ($-1$). The discrete solutions, which do not have continuous moduli parameters beyond the $\Omega$-deformation parameter, are represented by the black dots. Their $U(1)_R$ fluxes vanish.   The solution with $m=n=0$ is represented by a green dot: its $U(1)_R$ flux is $1$ ($-1$)  if  $A>0$ ($A<0$).

Let us briefly discuss how the solutions previously studied in the literature fit to our general classification.
For $n=-2$, the discrete solutions form a multiplet whose members are  labelled by  $m =-1,0,1$:
\bea
f \ \ &=& \ \ -1 \nn \\
\frac{\gamma^{(0)}}{\epsilon_{\Omega}} &=& \frac{m}{2} +\cos\theta
\nn \\
\frac{c_0}{\epsilon_{\Omega}} \ &=&\sqrt{\sin^2\theta +(\frac{m}{2}+\cos\theta)^2} \nn \\
G \ &=& \frac{ -\frac{m}{2} }{\sqrt{\sin^2\theta +(\frac{m}{2}+\cos\theta)^2}}\ . 
\eea

The solution found in \cite{Benini:2012ui} has
\bea
f = -1 \ , \qquad \gamma^{(0)} =\epsilon_{\Omega}\,\cos\theta \ , \qquad G =0
\eea
It  corresponds in our classification to the case with $m=0$ of the $n=-2$ multiplet~\footnote{Eq. (\ref{rank0}) shows that this solution has enhanced supersymmetry.}.

%%%%%%%%%%%%%%%%%%%%%%%%%%%%%%%%%%%%%%%%%%%%%%%%%%%%%%%%
\subsection{$\Omega$-Deformed $\mathbb{S}^2$}
\label{subsec:omegadeformed}

%%%%%%%%%%%%%%%%%%%%%%%%%%%%%%%%%%%%%%%%%%%%%%%%%%%%%%%%%

In this Section we will focus on the solution with $n=m=0$ (the green dot of Figure \ref{FluxesSphere}):
\bea
f = 0 \ , \qquad \gamma^{(0)} =\epsilon_{\Omega}\,A \ , \qquad G = \frac{\cos\theta  }{\sqrt{\sin^2\theta +A^2}} \ . 
\eea
$\mathcal{F}_R$ is given by Eq. (\ref{FRfinal}) and its flux 
  is $+1$ ($-1$) if $A>0$ ($A<0$)\footnote{The case with $n=0$ and $A=0$ is a singular limit: as discussed previously, if one sets $\gamma^{(0)}= 0$, the BRST transformations degenerate.}. In this subsection we show that this
background is  {\it topologically equivalent}, in the sense of Eq.  (\ref{topbkgrequivalence}),  to the so-called $\Omega$-deformed $\mathbb{S}^2$. The $\Omega$-deformed $A$-model was defined in  \cite{Closset:2014pda} and \cite{Closset:2015rna} by the equation
\bea
f= G. 
\label{ccbkgr}
\eea
We have already remarked that this equation implies the identification of the $\mathcal{F}_R$ with half of the
world-sheet curvature.  By substituting Eq. (\ref{ccbkgr})  into (\ref{dictionarytwo}), one obtains the equation
\bea
\frac{1}{2}\,\sqrt{g}\,\epsilon_{\mu\nu}\, D^\mu\,\gamma^\nu + f\, \gamma^{(0)}= f\,  \sqrt{\gamma^2 +(\gamma^{(0)})^2} \, . 
\eea
Taking account of (\ref{topgravitybkgs}) and (\ref{rounds2}), this gives
\bea
\epsilon_{\Omega}^2\,\cos\theta + \frac{\rmd\,\gamma^{(0)}}{\rmd \theta}\, \frac{\gamma^{(0)}}{\sin\theta}= \frac{\rmd\,\gamma^{(0)}}{\rmd \theta}\,\sqrt{\epsilon_{\Omega}^2 +\frac{(\gamma^{(0)})^2}{\sin^2\theta}} \ , 
\eea
which can be easily solved to yield
\bea
&& \frac{\gamma^{(0)}(\theta)}{\epsilon_{\Omega}} = -\frac{1}{2\, B} + \frac{B}{2}\, \sin^2\theta =  -\frac{1}{2\, B} + \frac{B}{2}\, V^2\nn\\
&& f= G =  B \, \cos\theta = \frac{B}{2}\, \sqrt{g}\, \epsilon_{\mu\nu}\, D^\mu \,V^\nu \nn\\
&& \frac{c_0}{\epsilon_{\Omega}}= \frac{1}{2\, B} + \frac{B}{2}\, V^2 \ . 
\label{omegabkg}
\eea
Since
\bea
\gamma^{(0)} =  -\frac{\epsilon_{\Omega}}{2\, B} + i_\gamma(\omega)\qquad
\mbox{and} \qquad f^{(2)} = \rmd \, \omega
\nn
\eea
where
\bea
\omega =\frac{B}{2}\,  g_{\mu\nu}\, V^\nu\, dx^\mu , 
\nn
\eea 
one verifies that the $\Omega$-deformed background (\ref{omegabkg}) is  indeed topologically gauge equivalent to 
\bea
&& \gamma^{(0)} = -\frac{\epsilon_{\Omega}}{2\, B}\nn\\
&& f= 0 \, 
\label{omegabkgbis}
\eea
i.e. to the  background (\ref{newbkgr}) with $n=0$ and $A=  -\frac{1}{2\, B}$.

In our considerations so far, both $G$ and $f$ are taken real-valued. There actually exists another class of  consistent SG backgrounds for which both
$G$ and $f$ are purely imaginary-valued. Formally, these backgrounds can be obtained from our backgrounds by  analytically continuing our formulas to pure imaginary values of $A$. The background with $A=i$, for example, is the situation discussed in \cite{Doroud:2012xw}.  For this ``Wick-rotated'' backgrounds, the two-dimensional flux configurations of the backgrounds correspond to exchanging $n$ and $m$ in Figure \ref{FluxesSphere}. 

%%%%%%%%%%%%%%%%%%%%%%%%%%%%%%%%%%%%%%%%%%%%%%%%%%%%%%%%
\subsection{All Localizing Backgrounds on $\mathbb{T}^2$}

For $\Sigma = \mathbb{T}^2$, let us adopt the coordinates 
\bea
\rmd s^2 = \rmd \theta_1^2 + \rmd  \theta_2^2 \, . 
\label{flattorus2}
\eea
We choose the vector field $\gamma$ to be one of the two Killing vectors:
\bea
\gamma = \epsilon_{\Omega}\,\partial_{\theta_1} \, . 
\eea 
Up to topological  equivalences,  the general solution of (\ref{topgravitybkgs}) is given by
\bea
f = 0=G \ , \qquad \gamma^{(0)} =\epsilon_{\Omega}\, A \ , \qquad c_0=\epsilon_{\Omega}\, \sqrt{1+ A^2} \ . 
\label{newbkgrtorus}
\eea
We see that the allowed values for the background fields are considerably reduced compared to those for $\mathbb{S}^2$. This is because the first Chern class of the topological $U(1)$ gauge field must vanish.

%%%%%%%%%%%%%%%%%%%%%%%%%%%%%%%%%%%%%%%%%%%%%%%%%%%%%%%
\section{$O(1,1)$ Duality Symmetry}
\label{sec:lorentzian}

In this section we will show that some of  the duality automorphisms of the supersymmetry algebra act as solutions generating symmetries.
To see this, let us return to the Killing spinor equation
\bea
(D_\mu - i \mathcal{A}_\mu) \,\zeta = - \frac{i}{2} \,f \,\Gamma_\mu \zeta + \frac{i}{2}\, G \,\Gamma_\mu \Gamma_3 \zeta \ . 
\label{eq:killingspinorbis}
\eea
We see that this equation is invariant under the global $O(1,1; \mathbf{R})$ transformations
\bea
\label{eq:lorentzone}
\begin{bmatrix} f \\ G \end{bmatrix}  & \rightarrow & 
\begin{bmatrix} f' \\ G' \end{bmatrix} = \begin{bmatrix} \cosh \alpha & \sinh \alpha \\
\sinh \alpha & \cosh \alpha \end{bmatrix}
\begin{bmatrix} f \\ G \end{bmatrix} 
\nn \\
\zeta \ \ &\rightarrow&  \ \ \zeta^\prime \ \ \ = \ \ \ \mathrm{e}^{\frac{\alpha}{2}\, \Gamma_3}\, \zeta\nn\\
\mathcal{A}_\mu &\rightarrow&  \ \ A^\prime_\mu  \  = \ \ \  \mathcal{A}_\mu  \ .
\eea
Namely, under the $O(1,1; \mathbf{R})$, $(f, G)$ transforms as a vector, $\zeta$ transforms as a spinor, while ${\cal A}_\mu$ is a scalar. We shall refer to this continuous global $O(1,1; \mathbb{R})$ invariance as ``non-compact duality symmetr''. 

Under the same $O(1,1; \mathbf{R})$ duality transformation, the topological bilinears transform as
\bea
\label{eq:lorentztwo}
\begin{bmatrix} c_0 \\ \gamma^{(0)} \end{bmatrix}
& \rightarrow & \begin{bmatrix} c_0' \\ (\gamma^{(0)})' \end{bmatrix} = \begin{bmatrix}
\cosh \alpha & \sinh \alpha \\ \sinh \alpha & \cosh \alpha
\end{bmatrix} 
\begin{bmatrix} c_0 \\ \gamma^{(0)} \end{bmatrix}
\nn \\
\gamma^\mu \ \ \ &\rightarrow& \ \ \ (\gamma^\mu)^\prime \ \ \ =\ \ \ \gamma^\mu \, . 
\eea
The $O(1,1; \mathbf{R})$ duality transformation leave the equations for the spinor bilinears (\ref{eq:bilinears}) invariant and thus
it must act on the topological backgrounds as well. 
%Therefore they act on the space of topological backgrounds. 
However, it is important to observe that  the $O(1,1; \mathbf{R})$ duality symmetry is realized {\it non-linearly} on the space of  solutions of the equations for TG backgrounds as follows:
%\bea
%D^m \, \gamma^n + D^n \, \gamma^m =0 \ , \qquad d\,\gamma^{(0)} =- i_\gamma (f^{(2)}) \ , 
%\label{topgravitybkgsbis}
% \eea
\bea
\label{eq:lorentzthree}
f \ \ \ \ \ &\rightarrow& \ \ \ \ \ f^\prime \ \ \ =\ \cosh\alpha \, f + \sinh\alpha\, G[f,\gamma^{(0)},\gamma^\mu]\nn\\
\gamma^{(0)} \ \  &\rightarrow& \ \  (\gamma^{(0)})^\prime \ =\ \sinh\alpha \, c_0[\gamma^{(0)},\gamma^\mu]+ \cosh\alpha\,\, \gamma^{(0)} \nn\\
\gamma^\mu \ \ \ &\rightarrow& \ \ (\gamma^\mu)^\prime \ \ = \ \ \ \gamma^\mu \, \nn\\
\mathcal{F}_R[f,\gamma^{(0)},\gamma^\mu] &\rightarrow& \mathcal{F}_R[f^\prime,(\gamma^{(0)})^\prime,(\gamma^\mu)^\prime]=\mathcal{F}_R[f,\gamma^{(0)},\gamma^\mu] \, , 
\eea
where
\bea
c_0[\gamma^{(0)},\gamma^\mu] \ \ &=& \sqrt{\gamma^2+(\gamma^{(0)})^2}\nn\\
G[f,\gamma^{(0)},\gamma^\mu] \ &=& \frac{1}{c_0} \left[ { \frac{1}{2}\,\sqrt{g}\,\epsilon_{\mu\nu}\, D^\mu\,\gamma^\nu + f\, \gamma^{(0)}} \right] 
%{ \sqrt{\gamma^2 +(\gamma^{(0)})^2}}
\nn\\
\mathcal{F}_R[f,\gamma^{(0)},\gamma^\mu] &=& \pm\,\frac{1}{2}\, \sqrt{(f^2- G^2-R)^2 + (D_\mu f)^2 - (D_\mu G)^2} \, . 
\label{compositetopologicalsugra}
\eea

On a compact surface $\Sigma$, the fluxes of $f$ and $G$ must be quantized. In general,  given a background configuration of (\ref{eq:killingspinorbis}) with quantized fluxes, the configuration obtained after the duality transformations (\ref{eq:lorentzone}) may not have quantized fluxes. When this happens, the transformed background is not physically acceptable. This implies that, for compact manifolds, the {\it continuous} duality symmetry $O(1, 1; \mathbf{R})$ is generically broken. 

Still, the theory might be invariant under a {\it discrete} set of duality transformations which send a configuration with quantized fluxes to another configuration with quantized fluxes: if $(n,m)$ are the $(f,G)$ fluxes of a given configuration, there must exists a nontrivial discrete duality transformation for each nontrivial integer fluxes  $(n^{\prime}, m^{\prime})$ that preserves the $O(1,1; \mathbf{Z})$ quadratic form
\bea
||(n', m')||^2 \equiv n^{\prime\; 2} - m^{\prime\; 2} = n^2- m^2 \equiv ||(n, m)||^2 \, . 
\label{discretelorentz}
\eea
As we already classified all solutions up to topological gauge equivalence, we can analyze the fate of the global duality symmetry in full generality. 

Take first the solutions with $|A|>\frac{|n|}{2}>0$ which have $f$ and $G$ fluxes equal to $(n,m)=(n,\pm n)$. We will focus on the  $n>0$ and $m=n$ class of background configurations, as the foregoing analysis would similarly hold for other classes. From (\ref{discretelorentz}), we see that the duality transformations which act on such backgrounds form a subgroup isomorphic to $\mathbb{Z}$, whose elements are the matrices for which 
\bea
{\mathrm e}^{\alpha_k} = k, \qquad \mbox{equivalently}, \qquad \alpha_k = \log k + 2 \pi i \mathbb{Z}
\eea
for some $k$ {\it positive} integer such that
\bea
\label{eq:lorentzdiscretek}
\begin{bmatrix} f\\ G\end{bmatrix} \ \ &\rightarrow& \ \ \begin{bmatrix} f^\prime \\ G^\prime \end{bmatrix}  \ \ =   \begin{bmatrix} \frac{1}{2} (k + \frac{1}{k}) &  \frac{1}{2} (k - \frac{1}{k})\\
 \frac{1}{2} (k - \frac{1}{k}) &  \frac{1}{2} (k + \frac{1}{k}) \end{bmatrix}\, 
\ 
\begin{bmatrix} f\\ G\end{bmatrix}\nn\\
\nn \\
\begin{bmatrix} \gamma^{(0)}\\ c_0\end{bmatrix} &\rightarrow& 
\begin{bmatrix} (\gamma^{(0)})^\prime \\ c_0^\prime \end{bmatrix}  
= 
\begin{bmatrix} \frac{1}{2} (k + \frac{1}{k}) &  \frac{1}{2} (k - \frac{1}{k})\\
 \frac{1}{2} (k - \frac{1}{k}) &  \frac{1}{2} (k + \frac{1}{k}) 
\end{bmatrix}\,
\begin{bmatrix}\gamma^{(0)}\\c_0\end{bmatrix} \, . 
\nn
\eea
Under such discrete duality transformation, the moduli $A$ and $n$ of the backgrounds are transformed to
\bea
\begin{bmatrix} n \\ A \end{bmatrix} \quad
\rightarrow \quad
\begin{bmatrix} n' \\ A' \end{bmatrix}
= \begin{bmatrix} 1 & 0 \\ 1/2 & 1 \end{bmatrix}
\begin{bmatrix} k & 0 \\ - k/2 & k \end{bmatrix}
\begin{bmatrix} n \\ A \end{bmatrix}  = 
k \begin{bmatrix} n \\ A \end{bmatrix}\ .  
\nn
\eea
We see that, starting from the solutions with $n=1$ and all  $A>1/2$, one can generate all other solutions with $n>0$ and  $A> n/2$ by $O(1,1; \mathbf{R})$ duality transformations.

On the other hand, generic discrete solutions for $A= \frac{m}{2}$ and fluxes $(n, m)$ with $|m| < |n|$ and $n\not=0$ breaks completely the $O(1,1;\mathbf{R})$ duality symmetry group (\ref{eq:lorentzone}). This is the case, for example,  of the solution $(n,m)=(-2,0)$ of \cite{Benini:2012ui},
since the only solution of
\bea
 n^{\prime\; 2} - m^{\prime\; 2} =4
\label{discretelorentz01}
\eea
are $n^\prime=\pm 2$ and $m'=0$.  In general, one can show that the set of $O(1,1; \mathbf{Z})$ duality transformations which send a given discrete solution into another discrete solution is a finite set (generically empty).  For example, the only other solution which can be generated by the Lorentzian symmetry from the  discrete solution $(n,m)=(7,2)$ is the one with $(n^\prime, m^\prime)= (9,6)$. 

It remains to consider the solution with $n=0$ and $A>0$ continuous\footnote{The solution with $A<0$ corresponds to the $\bar{A}$-twisted model.} associated with the $\Omega$-deformed $\mathbb{S}^2$
\bea
\gamma^{(0)} \ \ \ &=& \ \ \ \epsilon_{\Omega}\,A\nn\\
f \ \ \ \ \ &=& \ \ \  0\nn\\
c_0(\gamma^{(0)},\gamma^\mu) \ \ &=& \epsilon_{\Omega}\,\sqrt{\sin^2\theta+A^2} \nn \\
G(f,\gamma^{(0)},\gamma^\mu) &=& \frac{\cos\theta  }{\sqrt{\sin^2\theta +A^2}}
\label{omegabkgquater}
\eea
By a general $O(1,1; \mathbf{R})$ rotation, one obtains
\bea
&& \frac{(\gamma^{(0)})^\prime}{\epsilon_{\Omega}}=\sinh\alpha \,  \sqrt{\sin^2\theta+A^2}+ \cosh\alpha\, A 
\eea
and hence that
\bea
A^\prime= \,\frac{1}{2}\, \Big[ {( \gamma^{(0)})^\prime(\pi) + (\gamma^{(0)})^\prime(0)} \Big]
= \mathrm{e}^\alpha\, A \, . \nn
\eea
We see therefore that, starting from the $n=0$ $A=1$ model, one obtains all values $A>0$ by acting with the duality symmetry transformation. 
\begin{figure}
\centering
    \includegraphics[width=0.6\textwidth]{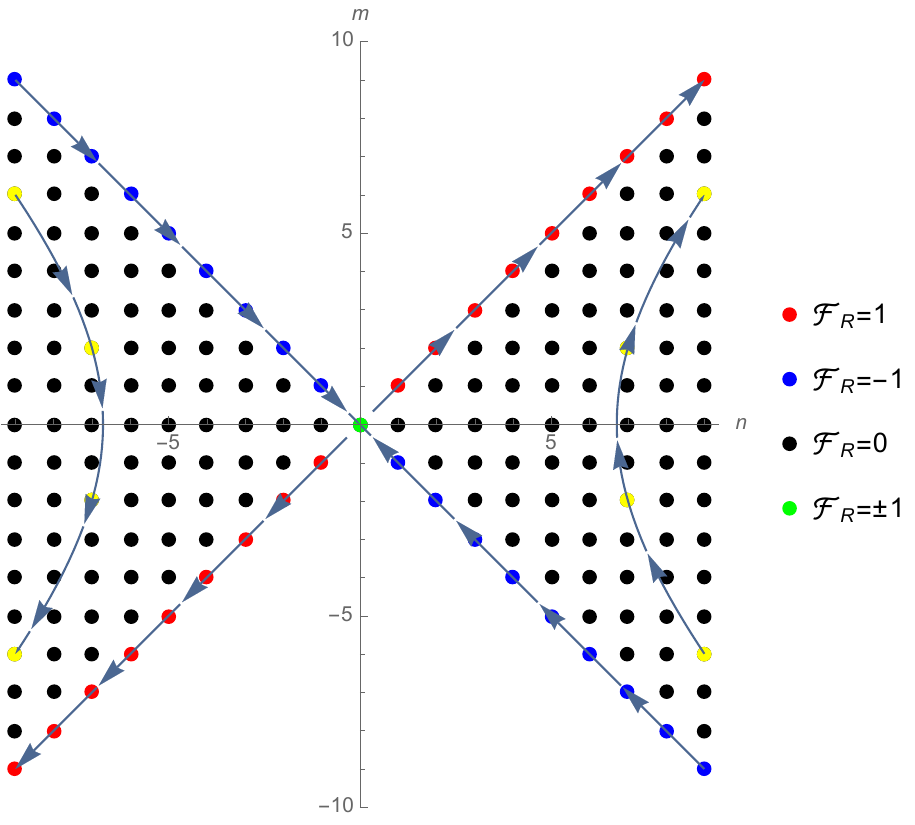}
   \caption{The action of the discrete non-compact duality symmetry on fluxes: $n=\int_{\Sigma} \sqrt{g}\, f$, $m=\int_{\Sigma} \sqrt{g}\, G$.}
    \label{DualityAction}
\end{figure}

Summarizing, the $O(1,1, \mathbb{R})$ duality  transformations generate, starting from the solutions with $n=\pm 1$ and $n=0$ in the continuous branch, all  other solutions with generic $n$ and $A$ continuous. On the other hand, the solutions in the discrete branch, with $A= \frac{m}{2}$ and $|m|<|n|$ are, generically,  not connected by the duality transformations. 
The action of the duality transformations on the localizing SG backgrounds is depicted in Figure \ref{DualityAction}.

%%%%%%%%%%%%%%%%%%%%%%%%%%%%%%%%%%%%%%%%%%%%%%%%%%%%%%%%
\section{Conclusions} 
\label{sec:conclusions}

In this work, we have obtained a complete classification of  supersymmetric localizing backgrounds (metric and gauge field) that can be constructed in two-dimensional $\mathcal{N}=(2,2)$ SG. The key idea has been to couple two-dimensional  matter TQFT to topological gravity and to relate BRST invariant topological backgrounds to supersymmetry preserving SG backgrounds. This approach was already  introduced in three dimensions in \cite{Imbimbo:2014pla}. The present work discusses its universality and generality and also explicitly works out the dictionary between the TG and SG approaches in two dimensions.

Unlike the  SG approach widely discussed in the literature, our analysis uses TG  to analyze and classify the manifolds admitting generalized covariantly constant spinors. In this paper, we showed that {\it all} the two-dimensional backgrounds which admit generalized covariantly constant spinors can be obtained via TG. More precisely, we demonstrated that there is a precise map which allows to reconstruct, given a BRST invariant topological background, a solution of the equations for generalized covariantly constant spinors in ${\cal N}= (2,2)$ SG. From a more technical point, we have learned that the two-dimensional case presents a new feature when compared to the three-dimensional one analyzed in \cite{Imbimbo:2014pla}: one
needs to introduce also a background $U(1)$  topological gauge field, equivariantly coupled to TG. 
 In Section \ref{sec:2dYM} we have shown that  the abelian gauge multiplet is necessary
to consistently couple the \ym2 theory  to TG. In other words, the rigid matter theory explicitly tells us what are the backgrounds  which need to be introduced. 

In fact, the topological approach also provides a natural and precise notion of {\it equivalent} backgrounds, i.e. backgrounds that can be made equal through  topological gauge transformations. In this way, it becomes much easier to obtain a complete classification of the localizing backgrounds. We believe this is a significant advantage over the more traditional SG approach, for which
the analogous notion has not yet been worked out. 

The natural implication of our map is the prediction  that the topological partition function $Z_{\rm top}[\gamma^\mu, f^{(2)}, \gamma^{(0)}]$ of the matter TQFT coupled to BRST invariant topological backgrounds is identical to the partition function of the matter SQFT coupled to the corresponding SG backgrounds
\bea
 Z_{\rm top}[\gamma^\mu, f^{(2)}, \gamma^{(0)}]= Z_{\rm sugra}\big[ f,  G[f,\gamma^{(0)},\gamma^\mu], \mathcal{F}_R[f,\gamma^{(0)},\gamma^\mu], \zeta  \big] \, . 
\eea
Here,  $\zeta$ is the covariantly constant spinor solution of Eq. (\ref{covariantlyconstantsugra}), while $G[f,\gamma^{(0)},\gamma^\mu]$ and $\mathcal{F}_R[f,\gamma^{(0)},\gamma^\mu]$ are  the ``composite'' SG backgrounds expressed in terms of the topological backgrounds in (\ref{compositetopologicalsugra}). Checking this prediction explicitly is an outstanding open problem
that we leave for the future.

We found many more supersymmetric localizing solutions than those that have been explored so far:  it would be interesting in particular to investigate the ``discrete'' solutions that generalize the one of \cite{Benini:2012ui} and \cite{Doroud:2012xw}.
In this regard, it would be also  interesting
to understand how the non-compact duality transformations act on the partition functions of different supersymmetric localizing backgrounds.

A direction for future investigation might be the uplift of the new supersymmetric localizing backgrounds we discovered to the superstring setup. The two-dimensional matter SQFTs with at least 2 supercharges contain vector, chiral and twisted chiral supermultiplets. Such systems arise as the low-energy limit of two-dimensional matter on intersecting D-branes. It could be interesting to study how the topological symmetry which we have uncovered is realized from the brane perspectives.

Another direction is to understand better the $\Omega$-deformation. We related this background to turning on the background of vector superghost in TG. It would be interesting to extend this to the most general $\Omega$-deformations and to utilize the two-dimensional analysis in this paper to the worldsheet formulation of the $\Omega$-deformations. 

As yet another direction, we note that the coupling of the matter TQFT to TG is also the starting point for constructing  topological strings. It seems reasonable to ask if the backgrounds with non-vanishing ghost number which were at the core of our analysis  are relevant in the topological string set up in which the  topological backgrounds become dynamical. One might speculate that the $\Omega$-deformation  be relevant to the world-sheet understanding of the refined topological string, in particular, in the context of computation of elliptic genera of M-strings in six dimensions 
\cite{Haghighat:2013gba}, \cite{Haghighat:2013tka}, \cite{Hohenegger:2013ala}, 
and monopole strings in five dimensions  \cite{Haghighat:2015coa}, \cite{Hohenegger:2015cba}.

We hope we made clear that the methods introduced in this paper are quite general and they are  not restricted either to two dimensions or to $\mathcal{N}= (2,2)$ SG:  we believe they might be a valuable tool to explore  localizing backgrounds also in other dimensions and with different supersymmetry content. For example, it should be relatively simple to study localizing backgrounds which arise in two-dimensional  $\mathcal{N}= (4,4)$ SG~\footnote{Some preliminary results for this case have been recently obtained in \cite{Lawrence:2015qra}.}.

%%%%%%%%%%%%%%%%%%%%%%%%%%%%%%%%%%%%%%%%%%%%%%%%%%%%%%%%%%
\section*{Acknowledgments}
We acknowlegde Dongsu Bak, Laurent Baulieu, Stefano Cremonesi, Andreas Gustavsson, Stefan Hohenegger, Euihun Joung, Seok Kim, Daniel Krefl, Diego Rodriguez-Gomez, Yolanda Lozano, Mauricio Romo, Eric Sharpe, Alessandro Tomasiello, Anderson Trimm, Cumrun Vafa, Alberto Zaffaroni and Yang Zhou for discussions. 
CI thanks the String Theory Group of the Seoul National University, DR thanks Theory Group at SISSA, SJR thanks the High-Energy Theory Group at Weizmann Institute of Science, the Institute for Theoretical Physics at University of Madrid, String Theory Group at Oviedo University and "Physics at the Riviera 2015" conference at Sestri Levante for their respective warm hospitality during this work. 
JB, SJR, DR acknowledges excellent collaboration environment provided by APCTP Focus Program "Liouville, Integrability and Branes (11)", where part of this work was done.  
The work of CI was supported in part by  INFN, by Genoa University Research Projects (P.R.A.) 2014 and 2015. The work of JB, DR, SJR was supported in part by the National Research Foundation of Korea grants 2005-0093843, 2010-220-C00003 and 2012K2A1A9055280.

\newpage

\appendix

\section{Standard YM Theory from Topological YM Theory } 
\label{App:AppendixA}
We asserted that, in two-dimensional spacetime, standard YM theory is related to topological YM theory by a certain deformation. Here, we explain details of this relation. The starting point is the partition function of the standard YM theory, viewed as a deformation of the topological YM theory:
\bea
Z[\epsilon] = \int [\rmd A\, \rmd \phi]\, \mathrm{e}^{- I_{\rm YM}[A, \phi]}=  \int [\rmd A\, \rmd \phi]\, \mathrm{e}^{- I_{\rm top}[A, \phi]}\, \mathrm{e}^{ - \epsilon \int_{\Sigma} \rmd^2x\,\sqrt{g}\,\frac{1}{2}\, \Tr\, \phi^2 } \ ,
\label{Zepsilon}
\eea
Here, $\epsilon$ is a positive semidefinite deformation parameter and
\bea
I_{\rm top}[A, \phi] = \int_{\Sigma}  \Tr\,\phi\, F
\eea
is the topological YM theory action. Expanded in power series of the deformation parameter $\epsilon$,
\bea
 Z_{\rm pert}[\epsilon] &= & \sum_{n=0}^\infty \frac{\epsilon^n}{n!}\int [\rmd A\, \rmd \phi]\, \mathrm{e}^{- \Gamma_{top}[A, \phi]}\,  \int_{\Sigma} d^2x_1\,\sqrt{g}\, \Tr\, \phi^2 (x_1)\, \cdots \int_{\Sigma} d^2x_n\,\sqrt{g}\, \Tr\, \phi^2 (x_n)\nn\\
\qquad &=&\sum_{n=0}^\infty \frac{\epsilon^n}{n!} \int_{\Sigma} d^2x_1\,\sqrt{g(x_1)}\cdots \int_{\Sigma} d^2x_n\,\sqrt{g(x_n)}\, \Big \langle\Tr\, \phi^2 (x_1)\cdots \Tr\, \phi^2 (x_n)\Big \rangle_{\rm top} \ ,
\nn
\\
\label{Zpert}
\eea
where $\langle\cdots\rangle_{\rm top}$ denotes the vacuum expectation value computed in the {\it topological YM theory}. 
Since we are expanding in $\epsilon$, $Z_{\rm pert}(\epsilon)$ is of course only the {\it perturbative} part of the
full partition function $Z(\epsilon)$ of the standard theory:  $Z_{\rm pert}(\epsilon)$  does differ from  $Z(\epsilon)$ by exponentially small, nonperturbative terms of $O\bigl(\mathrm{e}^{-\frac{1}{\epsilon}}\bigr)$.

One also observes that the zero-form 
\bea
O^{(0)} =   \Tr\, \phi^2 \ ,
\eea
is BRST invariant
\bea
s\,O^{(0)}   =0 \ ,
\eea
and hence
\bea
\rmd\,  \Tr\, \phi^2 = s\, O^{(1)} \ ,
\eea
where
\bea
 O^{(1)} =  \Tr\,\phi\, \psi \ .
 \eea
 This implies that the correlation functions
 \bea
 \big\langle\Tr\, \phi^2 (x_1)\cdots \Tr\, \phi^2 (x_n)\rangle_{top} \equiv \langle \bigl(\Tr\, \phi^2\bigr)^n \big\rangle_{\rm top} \ ,
 \eea
 does not depend on the operator insertion locations $x_1,\ldots x_n$.  So, the perturbative partition function  (\ref{Zpert}) becomes 
 \bea
  Z_{\rm pert}(\epsilon) =  \sum_{n=0}^\infty \frac{\bigl(\mathrm{Vol(\Sigma)}\,\epsilon\bigr)^n}{n!}  \Large\langle\bigl(\Tr\, \phi^2\bigr)^n\Large\rangle_{\rm top} \ ,
\label{Zpertsmooth}
\eea
where $\mathrm{Vol}(\Sigma)$ is the area of the surface $\Sigma$ with the chosen background metric $g$.

When $\Sigma$ is a closed surface $\Sigma_h$ of genus $h$, the topological correlation function  which appear in the expansion (\ref{Zpertsmooth}) is reduced to
integrals over the moduli space $\mathcal{M}_h$ of flat gauge connections on $\Sigma_h$. More precisely, 
the ghost number-four BRST operator $\Tr\, \phi^2$ corresponds to a closed four-form $\Omega^{(4)}$  on $\mathcal{M}_h$:
\bea
\Tr\, \phi^2 \qquad \leftrightarrow \qquad \Omega^{(4)} \ ,
\eea
so the topological correlation functions which appear in (\ref{Zpertsmooth}) become
\bea
 \Big\langle\bigl(\Tr\, \phi^2\bigr)^n\ \Big\rangle_{\rm top} = \int_{\mathcal{M}_h} \, \bigl(\Omega^{(4)}\bigr)^n\, \mathrm{e}^{\omega_2} \ ,
 \label{topintegration}
 \eea
where $\omega_2$  is the natural symplectic two-form on $\mathcal{M}_h$, defined by
\bea
\omega_2(\delta A, \delta A) = \int_{\Sigma} \Tr\, \delta A\wedge \delta A \ .
\eea
The correlation functions (\ref{topintegration}) make it clear that the series expansion (\ref{Zpertsmooth}) terminates after a finite number of terms so that $Z_{pert}(\epsilon)$ is a polynomial in $\epsilon$ whose degree depends on the genus $h$ of $\Sigma_h$.

The integration in (\ref{topintegration}) is well-defined as long as the moduli space $\mathcal{M}_h$ is smooth and compact. Singularities of $\mathcal{M}_h$ are associated with reducible flat connections. Therefore, in the presence of reducible flat connections, $Z_{\rm pert}(\epsilon)$ does not admit an expansion in integer positive powers of $\epsilon$, as in (\ref{Zpertsmooth}), but it also includes  terms with fractional, possibly negative,  powers of $\epsilon$. In this case,  $Z_{\rm pert}(\epsilon)$  still encodes some topological information of  the moduli space $\mathcal{M}_h$,  but it is not related in any simple way to the standard intersection numbers of it.

%%%%%%%%%%%%%%%%%%%%%%%%%%%%%%%%%%%%%%%%%%%%%%%%%%%%%%%
\section{$\mathbb{S}^1$-Equivariant Cohomology}

\subsection{$\mathbb{S}^1$-Equivariant Cohomology for $\mathbb{S}^2$}
\label{app:sphereS1S2}
A generic invariant two-form $f^{(2)}$
can be decomposed as
\bea
f^{(2)}= c_1\, \sqrt{g}\,\frac{1}{2} \epsilon_{\mu\nu} \rmd x^\mu\, \rmd x^\nu + \rmd \,\theta^{(1)} \ , 
\label{sphereone}
\eea
where $c_1$ is a constant and $\theta^{(1)}$ satisfies
\bea
\mathcal{L}_\gamma\, \theta^{(1)}= \rmd\, \omega^{(0)}
\ . 
\label{thetaoneinvariant}
\eea
The positive-definite  inner product on the space of one-forms
\bea
\big\langle\omega^{(1)}, \widetilde{\omega}^{(1)}\big\rangle =\int_{\Sigma} \omega^{(1)}\ast \widetilde{\omega}^{(1)}= \int_{\Sigma} \rmd^2x \sqrt{g}\, \omega_\mu \, \widetilde{\omega}_\nu\, g^{\mu\nu}
\eea
built with the $\mathcal{L}_\gamma$-invariant metric $g_{\mu \nu}$ is $\mathcal{L}_\gamma$-invariant. So, 
$\theta^{(1)}$  admits the orthogonal decomposition:
\bea
\theta^{(1)} = \theta^{(1)}_0 + \theta^{(1)}_{\perp} \ , 
\label{orthdecvectors}
\eea
where 
\bea
\mathcal{L}_\gamma \, \theta^{(1)}_0=0 
\eea
and
\bea
\big\langle \theta^{(1)}_{\perp}, \theta^{(1)}_0 \big\rangle =0, \qquad \mbox{and} \qquad  \theta^{(1)}_{\perp}= \mathcal{L}_\gamma\, \omega^{(1)}
\eea
since the image of $\mathcal{L}_\gamma$ is orthogonal to the space of invariant forms.

There is a positive definite, $\mathcal{L}_\gamma$-invariant  inner product on the space of zero-forms
as well:
\bea
\big\langle\omega^{(0)}, \widetilde{\omega}^{(0)} \big\rangle = \int_{\Sigma} \rmd^2x \sqrt{g}\, \omega\, \widetilde{\omega}
\nn
\eea
and the corresponding orthogonal decomposition:
\bea
\omega^{(0)} = \omega^{(0)}_0 + \omega^{(0)}_{\perp} \ . 
\label{orthdecscalars}
\eea
Then,
\bea
\mathcal{L}_\gamma \, \theta^{(1)} = \mathcal{L}_\gamma \, \theta^{(1)}_{\perp}= \rmd\,\omega^{(0)}_0 + \rmd\,  \omega^{(0)}_{\perp} \ . 
\nn
\eea
This implies
\bea
\rmd\,\omega^{(0)}_0 =0
\nn
\eea
and hence
\bea
\mathcal{L}_\gamma \, \theta^{(1)} =  \rmd\,  \omega^{(0)}_{\perp}= \mathcal{L}_\gamma \, \rmd\, \theta^{(0)} \, . 
\nn
\eea
In other words, 
\bea
\mathcal{L}_\gamma \, \widetilde{\theta} ^{(1)}\equiv \mathcal{L}_\gamma \,  (\theta^{(1)} - \rmd\, \theta^{(0)})=0 \, . 
\eea
Hence,
\bea
&& f^{(2)}= c_1\, \sqrt{g}\,\frac{1}{2} \epsilon_{\mu\nu} \rmd x^\mu \, \rmd x^\nu + \rmd \,\widetilde{\theta}^{(1)}\nn\\
&& \gamma^{(0)} = c_1\,  \gamma^{(0)}_{BC}+   i_\gamma(\widetilde{\theta}^{(1)}) + c_0
\eea
with $\widetilde{\theta} ^{(1)}$ which is $\mathcal L_\gamma$ invariant.

\subsection{$\mathbb{S}^1$-Equivariant Cohomology for $\mathbb{T}^2$}
\label{subsec:torus}
Again, a generic {\it invariant} two-form $f^{(2)}$
can be decomposed as
\bea
f^{(2)}= c_1\, \sqrt{g}\,\frac{1}{2} \epsilon_{\mu\nu} 
\rmd x^\mu \, \rmd x^\nu + \rmd \,\theta^{(1)} \ , 
\eea
where
\bea
\mathcal{L}_\gamma\, \rmd \, \theta^{(1)} = 0 = \rmd\, \mathcal{L}_\gamma\, \theta^{(1)} \, . 
\eea
Hence,
\bea
\mathcal{L}_\gamma\, \theta^{(1)} = \rmd \, \omega^{(0)} + h^{(1)} \, , 
\eea
where $h^{(1)}$ is harmonic:
\bea
\rmd \, h^{(1)}= \rmd^\dagger\, h^{(1)}=0. 
\eea
Using the same orthogonal decomposition as in (\ref{orthdecvectors}) for $\theta^{(1)}$, $\omega^{(0)}$ and $h^{(1)}$, we now obtain
\bea
\mathcal{L}_\gamma \, \theta^{(1)} = \mathcal{L}_\gamma \, \theta^{(1)}_{\perp}= \rmd \,\omega^{(0)}_0 + \rmd \,  \omega^{(0)}_{\perp}+ h^{(1)}_0+ h^{(1)}_{\perp} \ . 
\eea
From this,
\bea
\rmd \,\omega^{(0)}_0 + h^{(1)}_0=0
\eea
and hence
\bea
\mathcal{L}_\gamma \, \theta^{(1)} =  \rmd \,  \omega^{(0)}_{\perp}+  h^{(1)}_{\perp}= \mathcal{L}_\gamma \,(\rmd\, \theta^{(0)}-\widetilde{h}^{(1)}). 
 \eea
So, 
\bea
\widetilde{\theta}^{(1)} \equiv \theta^{(1)}-  \rmd\,\theta^{(0)}+\widetilde{h}^{(1)}
\eea 
is invariant: 
\bea
\mathcal{L}_\gamma \widetilde{\theta}^{(1)}=0
\eea
and $f^{(2)}$ is reduced to
\bea
&& f^{(2)}= c_1\, \sqrt{g}\,\frac{1}{2} \epsilon_{\mu\nu} \rmd x^\mu\, \rmd x^\nu +\rmd\, \widetilde{\theta}^{(1)} \, . 
%&&\gamma^{(0)}=  c_1\, \gamma^{(0)}_{BC}+ i_\gamma(\tilde{\theta}^{(1)}_0) +c_0 
\eea
However,  there is no nontrivial $U(1)$ bundle on $\mathbb{T}^2$ invariant under $\gamma$. We must therefore set
\bea
c_1=0 \, . 
\eea
This leads to
\bea
\gamma^{(0)}=  i_\gamma(\widetilde{\theta}^{(1)}) +c_0 \ .
\eea
%\subsection{Torus}
%There are no non-trivial line bundle on the torus which are $\gamma$-invariant.
%Therefore we must take $c_1=0$.  Then the most general solution is
%\bea
%&& f^{(2)}= d\, \theta^{(1)}_0\nn\\
%&&\gamma^{(0)} = c_0 + i_\gamma(\theta^{(1)}_0) 
%\eea
%The only modulus is  the constant $c_0$. 

\newpage
 %\begin{thebibliography}{99}
\providecommand{\href}[2]{#2}


\begin{thebibliography}{10}
%[\href{}{{\tt 	}}].
%\cite{Pestun:2007rz}
\bibitem{Pestun:2007rz} 
  V.~Pestun,
  {\sl Localization of gauge theory on a four-sphere and supersymmetric Wilson loops}, 
  Commun.\ Math.\ Phys.\  {\bf 313}, 71 (2012)
  %%CITATION = ARXIV:0712.2824;%%
  %405 citations counted in INSPIRE as of 05 Aug 2014
[\href{http://arxiv.org/abs/0712.2824}{{\tt arXiv:0712.2824}}].

%\cite{Rey:2010ry}
%\bibitem{Rey:2010ry}
  %S.~J.~Rey and T.~Suyama,
  %``Exact Results and Holography of Wilson Loops in N=2 Superconformal (Quiver) Gauge Theories,''
  %JHEP {\bf 1101} (2011) 136
  %[arXiv:1001.0016 [hep-th]].
  %%CITATION = ARXIV:1001.0016;%%

%\cite{Kapustin:2009kz}
%\bibitem{Kapustin:2009kz}
 % A.~Kapustin, B.~Willett and I.~Yaakov,
  %``Exact Results for Wilson Loops in Superconformal Chern-Simons Theories with Matter,''
  %JHEP {\bf 1003} (2010) 089
  %[arXiv:0909.4559 [hep-th]].
  %%CITATION = ARXIV:0909.4559;%%

%\cite{Kallen:2012cs}
%\bibitem{Kallen:2012cs}
  %J.~Källén and M.~Zabzine,
  %``Twisted supersymmetric 5D Yang-Mills theory and contact geometry,''
  %JHEP {\bf 1205} (2012) 125
  %[arXiv:1202.1956 [hep-th]].
  %%CITATION = ARXIV:1202.1956;%%

%\cite{Festuccia:2011ws}
\bibitem{Festuccia:2011ws} 
  G.~Festuccia and N.~Seiberg,
{\sl Rigid Supersymmetric Theories in Curved Superspace},
  JHEP {\bf 1106}, 114 (2011)
 % [arXiv:1105.0689 [hep-th]].
  %%CITATION = ARXIV:1105.0689;%%
  %158 citations counted in INSPIRE as of 31 Oct 2014 
  [\href{http://arxiv.org/abs/1105.0689v2}{{\tt arXiv:1105.0689}}].

%\cite{Karlhede:1988ax}
%\bibitem{Karlhede:1988ax}
 % A.~Karlhede and M.~Rocek,
  %``Topological Quantum Field Theory and $N=2$ Conformal Supergravity,''
 % Phys.\ Lett.\ B {\bf 212} (1988) 51.
  %%CITATION = PHLTA,B212,51;%%

%\cite{Imbimbo:2014pla}
\bibitem{Imbimbo:2014pla} 
  C.~Imbimbo and D.~Rosa,
{\sl Topological anomalies for Seifert 3-manifolds},
 % arXiv:1411.6635 [hep-th].
  %%CITATION = ARXIV:1411.6635;%%
  %2 citations counted in INSPIRE as of 14 juil. 2015
  [\href{http://arxiv.org/abs/1411.6635v2}{{\tt arXiv:1411.6635	}}].
 

%\cite{Johansen:1994aw}
%\bibitem{Johansen:1994aw}
  %A.~Johansen,
  %``Twisting of $N=1$ SUSY gauge theories and heterotic topological theories,''
  %Int.\ J.\ Mod.\ Phys.\ A {\bf 10} (1995) 4325
  %[hep-th/9403017].
  %%CITATION = HEP-TH/9403017;%%

%\cite{Rodriguez-Gomez:2014eza}
%\bibitem{Rodriguez-Gomez:2014eza}
 % D.~Rodriguez-Gomez and J.~Schmude,
  %``Partition functions for equivariantly twisted $ \mathcal{N}=2 $ gauge theories on toric Kähler manifolds,''
  %JHEP {\bf 1505} (2015) 111
  %[arXiv:1412.4407 [hep-th]].
  %%CITATION = ARXIV:1412.4407;%%


%\cite{Benini:2012ui}
\bibitem{Benini:2012ui} 
  F.~Benini and S.~Cremonesi,
{\sl Partition Functions of ${\mathcal{N}=(2,2)}$ Gauge Theories on S$^{2}$ and Vortices},
  Commun.\ Math.\ Phys.\  {\bf 334}, 1483 (2015)
 % [arXiv:1206.2356 [hep-th]].
  %%CITATION = ARXIV:1206.2356;%%
  %108 citations counted in INSPIRE as of 24 juil. 2015
  [\href{http://arxiv.org/abs/1206.2356v3}{{\tt arXiv:1206.2356}}].

%\cite{Doroud:2012xw}
\bibitem{Doroud:2012xw} 
  N.~Doroud, J.~Gomis, B.~Le Floch and S.~Lee,
{\sl Exact Results in D=2 Supersymmetric Gauge Theories},
  JHEP {\bf 1305}, 093 (2013)
 % [arXiv:1206.2606 [hep-th]].
  %%CITATION = ARXIV:1206.2606;%%
  %106 citations counted in INSPIRE as of 24 juil. 2015
  [\href{http://arxiv.org/abs/1206.2606v3}{{\tt arXiv:1206.2606}}].

 
%\cite{Closset:2014pda}
\bibitem{Closset:2014pda} 
  C.~Closset and S.~Cremonesi,
{\sl Comments on $ \mathcal{N} $ = (2, 2) supersymmetry on two-manifolds},
  JHEP {\bf 1407}, 075 (2014)
%  [arXiv:1404.2636 [hep-th]].
  %%CITATION = ARXIV:1404.2636;%%
  %11 citations counted in INSPIRE as of 14 juil. 2015
  [\href{http://arxiv.org/abs/1404.2636v3}{{\tt arXiv:1404.2636}}].
  
  %\cite{Closset:2015rna}
 \bibitem{Closset:2015rna} 
C.~Closset, S.~Cremonesi and D.~S.~Park,
{\sl The equivariant A-twist and gauged linear sigma models on the two-sphere}, 
  JHEP {\bf 1506}, 076 (2015)
 % [arXiv:1504.06308 [hep-th]].
  %%CITATION = ARXIV:1504.06308;%%
  %4 citations counted in INSPIRE as of 25 Jul 2015
  [\href{http://arxiv.org/abs/1504.06308v2}{{\tt arXiv:1504.06308}}].

%\cite{Witten:1992xu}
\bibitem{Witten:1992xu}
  E.~Witten,
{\sl 2-dimensional gauge theories revisited}, 
  J.\ Geom.\ Phys.\  {\bf 9} (1992) 303
%  [hep-th/9204083].
  %%CITATION = HEP-TH/9204083;%%
  %265 citations counted in INSPIRE as of 27 Feb 2014
[\href{http://xxx.lanl.gov/abs/hep-th/9204083}{{\tt arXiv:hep-th/9204083}}].

%\cite{Witten:1993yc}
\bibitem{Witten:1993yc} 
  E.~Witten,
{\sl  Phases of N=2 theories in two-dimensions,}
  Nucl.\ Phys.\ B {\bf 403}, 159 (1993) 
  [\href{http://xxx.lanl.gov/abs/hep-th/9301042v3}{{\tt arXiv:hep-th/9301042v3}}].
  %%CITATION = doi:10.1016/0550-3213(93)90033-L;%%
  %986 citations counted in INSPIRE as of 09 f�vr. 2016

%\cite{Witten:1991zz}
%\bibitem{Witten:1991zz} 
 % E.~Witten,
 % {\sl Mirror manifolds and topological field theory},
  %In *Yau, S.T. (ed.): Mirror symmetry I* 121-160
 % [hep-th/9112056].
  %%CITATION = HEP-TH/9112056;%%
  %314 citations counted in INSPIRE as of 20 janv. 2016
% [\href{http://xxx.lanl.gov/abs/hep-th/9112056}{{\tt arXiv:hep-th/9112056}}].

%\cite{Nekrasov:2002qd}
\bibitem{Nekrasov:2002qd} 
  N.~A.~Nekrasov,
 {\sl Seiberg-Witten prepotential from instanton counting},
  Adv.\ Theor.\ Math.\ Phys.\  {\bf 7}, no. 5, 831 (2003)
   [\href{http://xxx.lanl.gov/abs/hep-th/0206161}{{\tt arXiv:hep-th/0206161}}].
  %%CITATION = doi:10.4310/ATMP.2003.v7.n5.a4;%%
  %777 citations counted in INSPIRE as of 09 f�vr. 2016


%%\cite{tHooft:1974hx}
\bibitem{tHooft:1974hx} 
  G.~'t Hooft,
  {\sl A Two-dimensional model for mesons},
  Nucl.\ Phys.\ B {\bf 75} (1974) 461.
%%\cite{Migdal:1976}
\bibitem{Migdal:1976}
  A.~Migdal,
 {\sl Recursion equations in gauge field theories},
  Sov\ Phys.\ JETP {\bf 42}, (1976) 413.
%%

\bibitem{Baulieu:1988xs}
  L.~Baulieu and I.~M.~Singer,
 {\sl Topological Yang-Mills Symmetry}, 
  Nucl.\ Phys.\ Proc.\ Suppl.\  {\bf 5B} (1988) 12.

%\bibitem{Ouvry:1988mm}
  S.~Ouvry, R.~Stora and P.~van Baal,
{\sl On the Algebraic Characterization of Witten's Topological Yang-Mills
Theory},
  Phys.\ Lett.\  B {\bf 220} (1989) 159.

%\bibitem{Baulieu:1989rs}
  L.~Baulieu and I.~M.~Singer,
{\sl The Topological Sigma Model}, 
  Commun.\ Math.\ Phys.\  {\bf 125} (1989) 227.

% \bibitem{Kanno:1988wm}
  H.~Kanno,
{\sl Weyl Algebra Structure and Geometrical Meaning of BRST Transformation in Topological Quantum Field Theory},
  Z.\ Phys.\  C {\bf 43} (1989) 477.

%\cite{Gross:1993hu}
%\bibitem{Gross:1993hu}
  %D.~J.~Gross and W.~Taylor,
  %``Two-dimensional QCD is a string theory,''
 % Nucl.\ Phys.\ B {\bf 400} (1993) 181
  %[hep-th/9301068].
  %%CITATION = HEP-TH/9301068;%%

%\cite{Gross:1993yt}
%\bibitem{Gross:1993yt}
  %D.~J.~Gross and W.~Taylor,
  %``Twists and Wilson loops in the string theory of two-dimensional QCD,''
 % Nucl.\ Phys.\ B {\bf 403} (1993) 395
  %[hep-th/9303046].
  %%CITATION = HEP-TH/9303046;%%
  
%\cite{Imbimbo:2009dy}
\bibitem{Imbimbo:2009dy} 
  C.~Imbimbo,
{\sl The Coupling of Chern-Simons Theory to Topological Gravity},
  Nucl.\ Phys.\ B {\bf 825}, 366 (2010)
  [\href{http://arxiv.org/abs/0905.4631}{{\tt arXiv:0905.4631}}].
  %[arXiv:0905.4631 [hep-th]].
  %%CITATION = ARXIV:0905.4631;%%
  %3 citations counted in INSPIRE as of 01 Apr 2014

%\cite{Huang:2014gca}
%\bibitem{Huang:2014gca}
  %X.~Huang, S.~J.~Rey and Y.~Zhou,
  %``Three-dimensional SCFT on conic space as hologram of charged topological black hole,''
  %JHEP {\bf 1403} (2014) 127
  %[arXiv:1401.5421 [hep-th]].
  %%CITATION = ARXIV:1401.5421;%%
  %12 citations counted in INSPIRE as of 21 août 2015



%\cite{Gomis:2012wy}
\bibitem{Gomis:2012wy} 
  J.~Gomis and S.~Lee,
{\sl Exact Kahler Potential from Gauge Theory and Mirror Symmetry},
  JHEP {\bf 1304}, 019 (2013)
  %[arXiv:1210.6022 [hep-th]].
  %%CITATION = ARXIV:1210.6022;%%
  %60 citations counted in INSPIRE as of 08 Aug 2015
  [\href{http://arxiv.org/abs/1210.6022v2}{{\tt arXiv:1210.6022}}].

\bibitem{Haghighat:2013gba}
  B.~Haghighat, A.~Iqbal, C.~Kozcaz, G.~Lockhart and C.~Vafa,
{\sl M-Strings},
 [\href{http://arxiv.org/abs/1305.6322}{{\tt arXiv:1305.6322}}].

\bibitem{Haghighat:2013tka}
  B.~Haghighat, C.~Kozcaz, G.~Lockhart and C.~Vafa,
{\sl Orbifolds of M-strings},
  Phys.\ Rev.\ D {\bf 89}, no. 4, 046003 (2014)
   [\href{http://arxiv.org/abs/1310.1185}{{\tt arXiv:1310.1185}}].

\bibitem{Hohenegger:2013ala} S.~Hohenegger and A.~Iqbal, 
{\sl M-strings, elliptic genera and $\mathcal{N} = 4$ string amplitudes,} Fortsch.\ Phys.\  {\bf 62} (2014) 155 
 [\href{http://arxiv.org/abs/1310.1325}{{\tt arXiv:1310.1325}}].
 
 %%CITATION = ARXIV:1310.1325;%%

\bibitem{Haghighat:2015coa}
B.~Haghighat, {\sl From Strings in 6d to Strings in 5d,} 
 [\href{http://arxiv.org/abs/1502.06645}{{\tt arXiv:1502.06645}}].

%\cite{Hohenegger:2015cba}
\bibitem{Hohenegger:2015cba}
  S.~Hohenegger, A.~Iqbal and S.~J.~Rey,
  {\sl M-strings, Monopole strings and Modular Forms,}
  Phys.\ Rev.\ D {\bf 92} (2015) 6,  066005
   [\href{http://arxiv.org/abs/1503.06983}{{\tt arXiv:1502.06983}}].
  

%\cite{Lawrence:2015qra}
\bibitem{Lawrence:2015qra} 
  A.~Lawrence and M.~Soroush,
  {\sl N=(4,4) Vector Multiplets on Curved Two-Manifolds,}
  %arXiv:1509.00890 [hep-th].
  %%CITATION = ARXIV:1509.00890;%%
  %1 citations counted in INSPIRE as of 24 sept. 2015
 [\href{http://arxiv.org/abs/1509.00890}{{\tt arXiv:1509.00890}}].
\end{thebibliography}
\end{document}